\documentclass[pra,twocolumn,showpacs,preprintnumbers,amsmath,amssymb]{revtex4}

\usepackage{graphicx}
\usepackage{dcolumn}
\usepackage{subfigure}

\newcommand{\ket}[1]{|{#1}\rangle} \newcommand{\bra}[1]{\langle{#1}|}

\begin{document}

\title{Propagation of small fluctuations in electromagnetically induced
  transparency. Influence of Doppler
  width.}

\author{P. Barberis-Blostein} \affiliation{Instituto
  de Investigaciones en Matem\'aticas Aplicadas y en Sistemas. Universidad
  Nacional Aut\'onoma de M\'exico, Ciudad Universitaria, 04510 M\'exico D.F. M\'exico }
  \author{M. Bienert}%
\affiliation{Instituto de Ciencias F{\'\i}sicas,
  Universidad Nacional Aut\'onoma de M\'exico, Apartado Postal 48-3, 62251
  Cuernavaca, Morelos, M\'exico } 

\date{\today}

\begin{abstract}
  The propagation of a pair of quantized fields inside a medium of three-level
  atoms in $\Lambda$ configuration is analyzed. We calculate the stationary
  quadrature noise spectrum of the field after propagating through the medium
  in the case where the field has a general (but small) noise spectrum and the atoms are in a coherent population
  trapping state and show
  electromagnetically induced transparency (EIT). Although the mean values of
  the field remain unaltered as the field propagates,
  there is an oscillatory interchange
  of noise properties between the probe and pump fields. Also, as the field
  propagates, there is an oscillatory creation and annihilation of correlations
  between the probe and pump quadratures.  We further study the
  field propagation of squeezed states when there is two-photon resonance, but the field has a
  detuning $\delta$ from atomic resonance. We show that the field propagation is
  very sensitive to $\delta$. The propagation in this case can be explained as
  a combination of a frequency dependent rotation of maximum squeezed
  quadrature with an interchange of noise properties between pump and probe
  fields. It is also shown that the effect of the Doppler width in a squeezed state
  propagation is considerable.
\end{abstract}



\maketitle
\section{Introduction}

Electromagnetically induced transparency (EIT) \cite{rv:harris} emerges when coherence between electronic states of an atom suppresses the
absorption of incident light. An usually opaque medium consisting of such
atoms becomes transparent. Three electronic levels in a $\Lambda$-shaped
configuration are a paradigm for showing EIT when the two lower states are
coupled in two-photon resonance via the common excited state: In this case
destructive interference between the two excitation paths suppresses absorption
of photons. EIT has found many applications in coherent transfer of atoms
\cite{rv:transatomos}, laser cooling \cite{rv:lasercoolingteo,rv:lasercoolingexp}, and recently it was proposed to
serve as a quantum memory device for applications in quantum information
technology \cite{rv:memoria2,rv:lukincollo}.
\begin{figure}
  \includegraphics*[width=4cm]{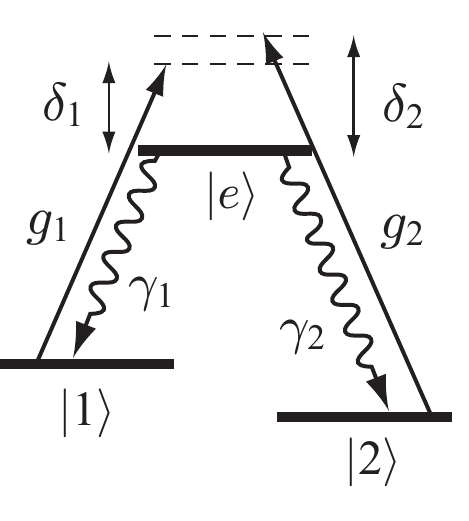}
  \caption{\label{fig:model} The atoms have a $\Lambda$ configuration with
    stable or metastable states $\ket 1$ and $\ket 2$ and common excited state
    $\ket e$. The transitions $\ket j\leftrightarrow \ket e$ with dipole
    coupling constants $g_j$ underlie spontaneous decay of rates $\gamma_j$
    $(j=1,2)$, the linewidth of $\ket e$ is $\gamma=\gamma_1+\gamma_2$.}
\end{figure}

The $\Lambda$-configuration is exemplified in Fig.~\ref{fig:model}, where two
(meta-)stable states, $|1\rangle$ and $|2\rangle$ are both coupled to an
excited state $|e\rangle$ by dipole interaction with the electromagnetic
fields of illuminating laser light. Spontaneous emission rates from the
excited state with linewidth $\gamma$ into $\ket i$ are denoted by $\gamma_i$ and
the detunings of the lasers' carrier frequency from the atomic transitions are
labeled by $\delta_i$. Detuned two-photon resonance is present if
$\delta_1=\delta_2=\delta\neq 0$.

\begin{figure}
  \includegraphics[width=6cm]{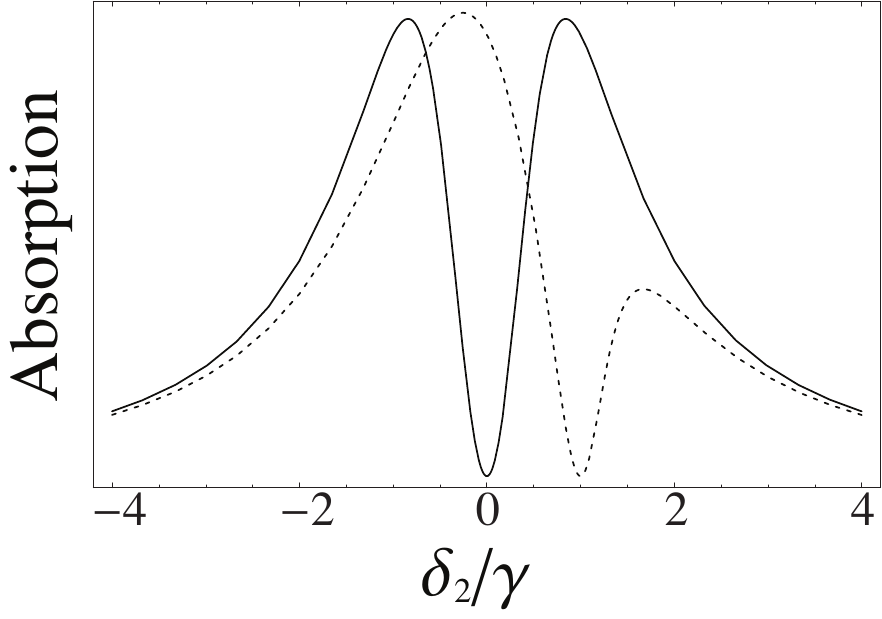}
  \caption{\label{fig:res} Absorption spectrum of the probe's mean value in
    the case of a resonant pump field, $\delta_1=0$, (solid line) and a
    detuned pump field $\delta_1=\gamma$ (dashed line), in arbitrary units.
    Here, $\delta_2=0$ refers to a resonant driving of the probe transition.
    Parameters: $\Omega_1=\Omega_2=\gamma.$}
\end{figure}

In the usual setup, the so-called pump laser drives one transition, e.g. $\ket
1\leftrightarrow\ket e$, while the probe laser, interacting with $\ket
2\leftrightarrow\ket e$, is tested for transparency \cite{lb:arimondo}.  The
Rabi frequencies associated with pump and probe are denoted by $\Omega_1$ and
$\Omega_2$. The linear response of the absorption of the probe by the medium
is described by the imaginary part of the electric susceptibility $\chi$ which is proportional to the mean value of the imaginary part of the electric
dipole.  In the case of two-photon resonance, $\delta_1=\delta_2$, the imaginary part of the
electric susceptibility $\chi$ vanishes and the medium becomes transparent for
the classical field. In Fig.~\ref{fig:res} we plot ${\rm Im}\chi$ as a
function of the probe detuning $\delta_2$ for the cases where the pump is in
resonance ($\delta_1=0$, solid line) and detuned ($\delta_1=\gamma$, dashed
line). The maximum absorption frequency of the probe field increases monotonically
with the Rabi frequencies.

There has been recent interest in understanding
the behavior of the quantum and noise properties of a field which propagates in EIT
media.  The studies concentrate on two cases: (i) when the strength of
the intensity of the pump field if much bigger than the probe field and (ii) when the intensity of the
pump field is of the same order as the probe field. 

For case (i), assuming a classical pump field and neglecting the
small absorption inside the EIT window, calculation shows that an incoming quantum state is the same
after propagation in the medium~\cite{rv:memoria2}. For the case where both
fields are quantized, with the resonant pump field being a coherent state and the incoming probe
state a squeezed vacuum, the absorption of squeezing from the different frequencies of a
broadband squeezed vacuum follows the classical EIT
window~\cite{rv:dantan4}. 
Different experiments that measure propagation of a
probe squeezing vacuum use this fact in
order to explain the propagated field ~\cite{rv:lvovsky2,rv:kozuma4} . 

When the absorption of the squeezed vacuum follows the classical EIT window,
we can write the effect of the medium on the field in the
following way~\cite{rv:lvovsky2}. Let $a(\omega)$ be the
annihilation operator for frequency $\omega$. Then, after propagation:
\begin{equation}\label{eq:naiveapp}
a''(\omega)=T(\omega) a(\omega)+\sqrt{1-|T(\omega)|^2}v(\omega)\, ,
\end{equation}
where $v(\omega)$ is the vacuum contribution and $T(\omega)$ is the
transmission function from the classical measurement.  

In some cases it is necessary to include the atomic noise generated by the atoms due to decoherence in the base
levels in order to explain the experimental results~\cite{rv:hetet1,rv:hetet2}.

For case (ii) a theoretical study of general phase noise propagating in
EIT was carried out in references~\cite{rv:fleich1,rv:richter}, in the
stationary regime. In these studies, the spectrum of the difference between the phase noise of both
fields is treated. They found that as the field propagates the phase noise of both fields
correlate and tend to be the same. Inside the validity domain of the approximation, 
the length of the formation of this
correlation follows the classical EIT transparency curve. The more transparent the medium
for the mean values, the longer the distance the field has to propagate in order
to get correlated.  Note that this result, although it does not give the
noise spectrum for each field, tells us that the field indeed changes as it
propagates and the scale of this change is given by the scale of the build-up
of the correlations.  Nevertheless, changing the field strength can make this
scale very long. This result is similar to that in
Ref.~\cite{rv:dantan4}, in the sense that the distance scale where the field
changes follows the classical EIT transparency curve.

There are also several experimental studies of the propagation of probe and pump field noise and
its correlations \cite{rv:lezama,rv:scullyeitco}.
They explain the results using the known effect of transformation
of the   incoming laser phase-noise to intensity noise. This transformation
happens due to small absorption by the atoms because of decoherence effects. In
the case of perfect EIT , this process would not exist, and the correlations
would not emerge.

Recent calculations, where both fields, probe and pump, are treated quantum
mechanically, show what happens in the case where the probe field is initially
a squeezed state.
It is shown, in the stationary regime and 
where there is no decoherence between the base levels, that
although the medium is transparent for the mean values of the field, initial quantum fluctuations are not
necessarily preserved after interaction. In
the case of a cavity filled with atoms in $\Lambda$ configuration driven by a
squeezed pump field and a coherent pump field, the quantum properties of the
output field can be very different from the quantum properties of the input
field \cite{rv:pablo,rv:pablo2}.  

Similar calculations were done in the case of a probe
squeezed state and a pump coherent state propagating in a medium showing EIT.
When the probe and pump carrier frequencies drive the atoms in resonance with their respective
transition, there is an oscillatory interchange of
noise properties between the initially squeezed probe field and the initially
coherent pump field as the state propagates
inside the medium.  The length of this interchange of squeezing can
be much smaller than the length where the phase noise of both fields correlate
~\cite{rv:richter} \cite{rv:pablomarc}. This means that the state measured after interacting with
the EIT medium can be totally different to the incoming state and that the
kind of analysis expressed by Eq.~\eqref{eq:naiveapp} is not always
valid. Note that this interchange of squeezing between probe and pump fields
is not related to absorption or noise generated by the atoms. 

In this paper, treating both fields quantum mechanically, we study (i) the spectrum of the probe and pump quadrature and
the correlations between them when the pump and probe field have some initial general
noise, (ii) the case where there is two-photon resonance, but the carrier
frequencies of the fields are
not in resonance with the corresponding atomic transitions ($\delta_1=\delta_2=\delta\neq
0$)
\footnote{In a previous publication \cite{rv:pablo2} we called this case two-photon
  detuning. As this term is confusing, we use the term of detuned two
  photon resonance for the case $\delta_1=\delta_2=\delta\neq 0$.}, 
 and (iii) the effect of the atoms' Doppler width in the field propagation. 
We find that the interchange of noise properties between squeezed states and
coherent states described
in~\cite{rv:pablomarc} extends to the case of general noise. Thus, for some
spectrum frequency and distance, the probe noise becomes equal to the initial
pump noise, and vice versa. Also, as the field propagates, there is an oscillatory creation and annihilation of correlations between the
pump and probe fields.
Our results allow, for example, to study the propagation of a probe squeezed
state with a pump laser with some phase noise. Although this is usually the
case, it has not been treated theoretically before, except for the case where only
difference in the phase noise was studied~\cite{rv:richter}. Our results
allows us to predict the quadrature of each field and the correlation between
them and not only the spectrum of the phase difference. Furthermore, we do not make an
adiabatic approximation: Our solution is valid for all frequencies.

We find also new qualitative behavior that appears
neither in the semiclassical nor in the resonance case.
If the carrier frequencies of the pump and probe fields drive the atoms with the same detuning from their
respective dipolar transition then, the mean value of the field stays
unaltered. Basically this means that as long as there
is two-photon resonance, the photon detuning from the atomic transition does not have a
strong effect in the propagation of classical pulses in EIT. 
Nevertheless, our analysis shows that a detuned (from the atom transition) two-photon resonance of the carrier
frequencies have a large impact on the propagation of the field state. 

Our
results show that the propagation of general states in EIT media is richer
than was usually believed. In particular we show that the maximum squeezed
quadrature of an initial broad squeezed vacuum rotates as it propagates in
the medium, the velocity of rotation depends on the detuning from the atom
transition and the spectrum frequency of interest.
This means that after some propagation, the
maximum squeezed quadrature of the propagated field is different for each
frequency. This spectrum-frequency velocity dependent rotation of the quadratures implies
that the output field is different to the incoming field for length scales
where the mean values of the field are almost unchanged. 

Note that the reduced velocity of pulse propagation in an EIT medium implies a
rotation of the maximum squeezed quadrature. This quadrature rotation is
due to the phase difference between the pulse, after propagating inside the
medium, with respect to the local oscillator used to measure the
quadratures. This phase difference depends on the pulse velocity. Nevertheless
this phase difference is global and because of that the quadratures for each
spectrum-frequency are rotated the same amount.
The fact that we obtain that the velocity of rotation depends on the spectrum
frequency makes this result qualitatively different from the quadrature
rotation due to slow light propagation.

We also show that in the
case where the probe and pump field have the same Rabi frequency, propagation is a
combination of the rotation of the squeezed quadrature plus interchange of
noise properties between the probe and pump field. 

We also present the influence of detuned two-photon resonances on EIT media with some
Doppler width.
The movement of the atoms in a medium causes a shift of the frequency that the
atoms ``see'' in their reference frame. Due to this Doppler effect, the atom
experiences a velocity-dependent frequency shift relative to the carrier
frequency of the fields measured in the laboratory frame.  For a thermal vapor
cell at room temperature, the Doppler width can be several times the decay
rate. If the pump and probe fields propagate in the same direction, then the
Doppler effect consists of detuning the frequency of both fields by the same
amount.  This implies that the propagation of the mean value of the field would be
negligibly disturbed, because, independently of the Doppler width, all the
atoms would be in two-photon resonance and the media shows EIT
\cite{lb:arimondo}.  This observation is one reason why
conventional (mean value measurement of the field) EIT experiments can be performed at room temperature.
In contrast, a detuned two-photon
resonance has a strong effect on the propagation of  quantum
states. This implies a significant impact of the atoms' Doppler width on the
quadrature spectra measured after propagation. 
There are many recent experiments where the propagation of vacuum squeezed
states is studied~\cite{rv:kozuma,rv:kozuma2,rv:kozuma3,rv:lvovsky1,rv:lvovsky2}. 
Our results impose a restriction on the possibility of conducting EIT experiments with
squeezed states in denser thermal clouds, even when we can neglect decoherence
between the base levels.

The paper is organized as follows: in section \ref{sc:equations} we state the
equations for the field and for the atoms, in
section \ref{sc:resonance} we study the case where the carrier frequencies of
the probe and pump field are in resonance and the incoming field has some
general noise, in section \ref{sc:detuning} we
study the case where the carrier frequencies of the probe and pump field are
in a detuned two-photon resonance ($\delta_1=\delta_2=\delta\neq 0$), in section \ref{sc:doppler}
we study the effect of the Doppler width on the propagation of the field.

\section{Theoretical description of the propagating fields}
\label{sc:equations}

The dynamics of the composite system of atoms and pump and probe  fields
are described by Heisenberg's equation of motion~\cite{rv:richter},
\begin{subequations}\label{eq:propagation}
\begin{alignat}{1}
  \left(\frac{\partial}{\partial t}+c\frac{\partial}{\partial z}\right)a_j =
  -i g_j N \sigma_{je}\, , 
  \label{eq:fprop}
\end{alignat}
and
\begin{alignat}{1}
  \frac{\partial}{\partial t} \varpi_1 =&\frac{1}{3}(-\gamma_1-\gamma)(1+ \varpi_{1}+\varpi_2)-2 i g_1(\sigma_{e1}a_1  - a^\dagger_1\sigma_{1e})\nonumber\\ & - i g_2(\sigma_{e2}a_2-a^\dagger_2\sigma_{2e})  + f_{\varpi_1},\nonumber\\
  \frac{\partial}{\partial t}\varpi_2 =&\frac{1}{3}(-\gamma_2-
  \gamma)(1+\varpi_{1}+\varpi_2)
  -i g_1(\sigma_{e1}a_1  -a^\dagger_1\sigma_{1e})\nonumber\\ & - 2 i g_2 (\sigma_{e2}a_2-a^\dagger_2\sigma_{2e}) +f_{\varpi_2},\nonumber\\
  \frac{\partial}{\partial t} \sigma_{1e}=&(-\frac{\gamma}{2}+i\delta_1)
  \sigma_{1e}+i g_1 \varpi_1 a_1-i g_2 \sigma_{12} a_2
  +f_{1e},\nonumber\\
  \frac{\partial}{\partial t} \sigma_{2e}=&(-\frac{\gamma}{2}+i\delta_2)
  \sigma_{2e}+i g_2 \varpi_2 a_2 -i g_1 \sigma_{21} a_1
  + f_{2e} ,\nonumber\\
  \frac{\partial}{\partial
    t}\sigma_{21}=&(-\gamma_{12}-i[\delta_1-\delta_2])\sigma_{21}-i g_1 a_1^\dagger
  \sigma_{2e} +i g_2 \sigma_{e1} a_2+f_{21}.
  \label{eq:aprop}
\end{alignat}
\end{subequations}
where $j=1,2$.
This description relies on a multi-mode decomposition of the
electromagnetic fields $\vec E_j=\vec{\mathcal E}_j a_j(z,t) \exp[ik_{j}z-i\omega_{j} t] + {\rm h.c.}$ around the carrier frequencies
$\omega_{j}=c k_{j}$, where $|\vec{\mathcal E}_j|$ is the
corresponding vacuum electric field. The detuning with respect to the atomic
transition $\omega_{je}$ is given by $\delta_i=\omega_i-\omega_{je}$. In this notation, the field envelope
operators $a_j(z,t)$ are slowly varying in space and time, allowing us to write
Maxwell's equation in the form \eqref{eq:fprop}, where the atomic polarization
proportional to $\sigma_{je}$ acts as a source.
  
The atomic operators $\sigma_{\mu\nu}(z) = \lim_{\Delta z\rightarrow 0}
\frac{L}{N\Delta z}\!\sum\limits_{z^{(j)}\in \Delta z }\!
\sigma_{\mu\nu}^{(j)}$ and $\varpi_j=\sigma_{ee}-\sigma_{jj}$ are written in the continuum limit, where
$\sigma^{(j)}_{\mu\nu}=\ket{\mu}^{(j)}\bra{\nu}$ is the atomic operator of
atom $j$ at position $z_j$. Here $N$ is the number of atoms, $L$ the length
of the medium and $\Delta z$ a space region around $z$. This approximation is justified if the
inter-atomic distance is smaller than the length scale introduced by the
wavelength of the carrier fields.

The coupling between atoms and field relies on a dipole interaction with
coupling constant $g_j=\vec \wp\cdot\vec{\mathcal E}_j/\hbar$, where $\vec \wp$ is
the atomic dipole moment. In order to arrive at the form of equations
\eqref{eq:fprop} and \eqref{eq:aprop}, the rotating wave approximation was
performed and all the operators are in a reference frame rotating with the
corresponding carrier frequencies.  The laboratory frame notation can be
obtained by the transformations $\sigma_{ej}'={\sigma}_{ej}\exp{[-i(
  \delta_{j}+\omega_{je})]}$ and $\sigma_{12}'={\sigma}_{12}\exp{[-i
  (\delta_{1}-\delta_{2})]}$.

To account for the noise introduced by the coupling of the atomic system to
the free radiation field, delta-correlated, collective Langevin operators,
$f_j$, were introduced. They have vanishing mean values and correlation
functions of the form $\langle f_{x}(z,t)
f_{y}(z,t')\rangle=\frac{L}{N}D_{xy}\delta(t-t')\delta(z-z')$. The diffusion
coefficients $D_{xy}$ can be obtained from the generalized Einstein equations
\cite{lb:louisell}. Their explicit form can be found, for example, in
\cite{rv:pablonicim}.

In order to obtain the spectrum of the quadrature fluctuations, the system of
equations is solved in the small-noise approximation using the standard technique of
transforming them into a system of c-number stochastic differential equations
\cite{rv:davidovich}.  The initial conditions for our analysis are:
The probe and pump field has $\langle a_j\rangle=\alpha_j$, thus driving the atomic transitions with Rabi
frequency $\Omega_j = |g_j\alpha_j|$.

Denoting the fluctuation of the $\theta$ quadrature of field $j=1,2$ as
\begin{equation}
  \delta Y_j^\theta(z,t)=\delta
  a_j(z,t) \exp(-i\theta) + \delta a_j^\dagger(z,t) \exp(i\theta)\, ,
\end{equation} 
where $\delta o=o-\langle o\rangle$, the $\theta$-quadrature noise
spectrum, in the stationary regime, is given
by
\begin{equation} {\mathcal S}_j(z,\omega) = \int\limits_{-\infty}^\infty
  e^{-i\omega t} \langle \delta Y_j^\theta(z,t)\delta Y_j^\theta(z,0)\rangle dt\, ,
 \label{eq:spec}
\end{equation}
and the correlations noise spectrum between, the probe $\theta_1$-quadrature
and the pump $\theta_2$-quadrature, is given by
\begin{equation} {\mathcal S}_c(z,\omega) = \int\limits_{-\infty}^\infty
  e^{-i\omega t} \langle \delta Y_1^{\theta_1}(z,t)\delta
  Y_2^{\theta_2}(z,0)\rangle dt\, ,
 \label{eq:cospec}
\end{equation}
where $\omega=0$ corresponds to the carrier frequency of the field in
accordance with the co-rotating reference frame we use.  The initial
conditions for the fluctuations can be written as


\begin{eqnarray}
  {\mathcal S}_1(z=0,\omega)&=&1+2 g_1(\omega) \cos(2 \theta)+2 f_1(\omega)\nonumber\, ,\\
  {\mathcal S}_2(z=0,\omega)& =&1+2 g_2(\omega) \cos(2 \theta)+2 f_2(\omega)\, ,\nonumber\\
{\mathcal S}_{c}(z=0,\omega)& =&0 \, , 
\label{eq:ics2}
\end{eqnarray}
where $f_j(\omega)$ and $g_j(\omega)$, $j=1,2$ are real. 


\section{Carrier frequencies in resonance with
  atomic transitions ($\delta_1=\delta_2=0$)}
\label{sc:resonance}
We solve equations~\eqref{eq:propagation} with
$\delta_1=\delta_2=0$ and noise spectrum given by~\eqref{eq:ics2}, following
the treatment used in \cite{rv:pablomarc}. We list in
appendix~\ref{ap:rescase} the analytical expressions for the quadrature noise
spectra of the  pump
($j=1$) and probe ($j=2$) field, and the correlations between them.

The behavior of the spectrum is characterized by two quantities: as the field
propagates $Q^{(i)}$ characterizes an exponential decay of noise properties and
$Q^{(r)}$ characterizes the coherent propagation of the field which corresponds
to an oscillatory evolution of noise properties. 

The behavior of $Q^{(i)}$ as a function of the parameters resembles the
behavior of the classical transmission spectrum ($T(\omega)$ in
Eq. \eqref{eq:naiveapp}). The absorption of the initial properties of the
field is then characterized by this quantity and allow us to define a length
scale, $z_{abs}$, where
these absorptions are important,
\begin{equation}\label{eq:absres} 
z_{\rm abs}\approx 1/(|Q^{(i)}|)\, .
\end{equation}

\subsection{Propagation of the field quadratures}
When $Q^{(i)} z\ll 1$, that is, when
we consider positions $z$ where the exponential absorption of the fluctuations
can be neglected, we can easily understand the effect of the oscillatory behavior
characterized by $Q^{(r)}$. Let us assume that $\alpha _1=\alpha _2$, then, as
the field propagates, there
is a complete oscillatory interchange of noise properties:
\begin{equation}\label{eq:interosc}
\mathcal{S}_2(Q^{(r)}z=k\pi)=\mathcal{S}_1(Q^{(r)}z=(k+1)\pi)\, ,
\end{equation}
\begin{equation}\label{eq:interosc2}
\mathcal{S}_j(Q^{(r)}z=k\pi+\pi/2)=(\mathcal{S}_1(z=0)+\mathcal{S}_2(z=0))/2\, ,
\end{equation}
where $k$ is an integer an $j=1,2$.

\begin{figure*}
  \subfigure{
    \includegraphics[width=16cm]{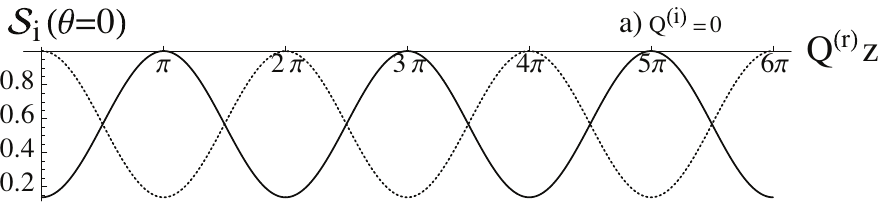}}\\
  \subfigure{\includegraphics[width=16cm]{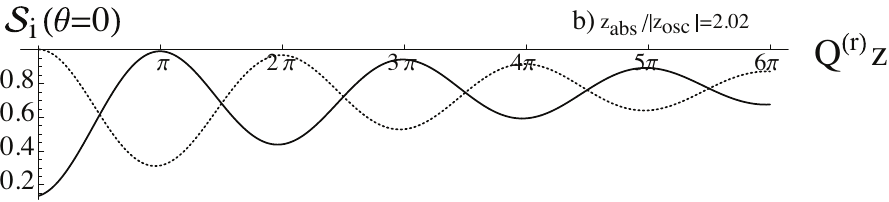}}
  \caption{\label{fig:mainres} Quantum properties of probe and pump fields propagating
    through an EIT medium. (a) The fluctuation spectrum of the probe
    (solid), initially (at $z=0)$ in a squeezed state with squeezing parameter
    $\xi=1$ and of the pump  (dashed), initially in a coherent state, as a
    function of position for $\theta=0$ and no absorption ($\gamma_i=0$). (b)
    Same as (a) but with absorption. The amplitude of the oscillations decays
    $1/e$ after $z_{\rm abs}/|z_{\rm osc}|=2.02$ periods.}
\end{figure*}

In Fig.~\ref{fig:mainres}a) we
have plotted the fluctuation spectrum for an initially squeezed probe (solid)
and an initially coherent pump (dashed), as a function of propagation
length $z$. For a
complete discussion when the probe field is a squeezed state, see
\cite{rv:pablomarc}.

Equations~\eqref{eq:interosc} and~\eqref{eq:interosc2} clearly display a
spectrum-frequency dependent oscillatory
transfer of the initial noise properties of the probe to the pump and back
while traveling through the medium, and it extends the result of
\cite{rv:pablomarc} to any initial noise.  The length scale of the oscillatory transfer
$z_{\rm osc}=2 \pi/Q^{(r)}$ can be much smaller than the absorption length
scale $z_{\rm abs}$. In fact
\[
\frac{z_{\rm abs}}{|z_{\rm
    osc}|}=\frac{|\Omega^2-\omega^2|}{\pi\gamma\omega}\, .
\]
In the case where $\omega\gamma\ll\Omega^2$ (corresponding to observations
frequencies inside the transparency window) we have $z_{\rm abs}\gg |z_{\rm
  osc}|$.

In Fig.~\ref{fig:mainres}b), the interplay of both scales, oscillatory and
absorption, can be clearly observed.

The oscillatory behavior implies that the outgoing field can be
completely different from the incoming field, although the mean values stay
exactly the same. This effect cannot be explained with the usual approach
expressed by Eq.\eqref{eq:naiveapp}. Moreover, it is qualitatively different from the
absorption since the noise properties, for each frequency, is recovered
after some propagation.

\subsection{Propagation of correlations}
A previous study of propagation of noise correlation between the pump
and probe field  was done in \cite{rv:richter}. They studied the
particular case of phase difference fluctuations.   
We first obtain a generalization (for the no-decoherence case) of the known
result of the spectrum of phase difference
fluctuations given in~\cite{rv:richter}. We present here the result for no decoherence in the base
level, but valid for all frequencies (no adiabatic approximation). Let
$\phi_j$ represent the phase of field $j$. We use the
fact that
for small noise the phase quadrature noise is proportional to the phase
noise~\cite{rv:davidovich}. 
From Eqs.~\eqref{eq:s1res}, ~\eqref{eq:s2res} and ~\eqref{eq:cores}, we obtain ($\alpha_1=\alpha_2=\alpha$):
\begin{eqnarray}\label{eq:comorichter}
\mathcal{S}_\phi(\omega,z)&=&(\delta\phi_1-\delta\phi_2)^2\nonumber\\&\approx&\frac{1}{\alpha^2}\{\mathcal{S}_1^{\theta=\pi/2}(\omega,z)+\mathcal{S}_2^{\theta=\pi/2}(\omega,z)\nonumber\\&&-2\mathcal{S}_c^{\theta_1=\theta_2=\pi/2}(\omega,z)\}\nonumber\\&=&\frac{1}{\alpha^2}e^{-{Q^{(i)}(\omega)}z}\mathcal{S}_\phi(\omega,z=0)\, .
\end{eqnarray}
Observe that as the field propagates, the difference between the phase noise of
the fields disappears, which implies that correlations between the phase noise of
both fields builds up. For the case of no decoherence, Eq. \eqref{eq:comorichter} is an extension of  Eq.(12) in
Ref.~\cite{rv:fleich1}, in the sense that it is valid for all $\omega$.
It is easy to see that the length of the fading of the phase
difference between the fields goes like $z_{abs}$, Eq.~\eqref{eq:absres}. For the phase noise difference, we do not observe any oscillatory behavior, as in the case
of the propagation of the field quadratures. 

Nevertheless,
the oscillatory transfer of initial noise properties between the pump and
probe field generates also an oscillatory creation and annihilation of
correlations (defined in Eq.~\eqref{eq:cospec}). When $z\ll z_{\rm abs}$, from Eq.~\eqref{eq:cores} we obtain
\begin{eqnarray}\label{eq:correso}
{\mathcal S}_{c}(z,\omega) &=&\sin \left(Q^{\text{(r)}} z\right) [\sin
\left(\theta _1-\theta _2\right) \left(f_2(\omega )-f_1(\omega
)\right)\nonumber\\&& +\sin
\left(\theta _1+\theta _2\right) \left(g_1(\omega )-g_2(\omega )\right)]\, .
\label{eq:lasteq}
\end{eqnarray}
Equation~\eqref{eq:lasteq} shows that if the noise properties of the pump and probe
field are different, then correlations between the fields  oscillates for length scales
 small with respect to the absorption scale. Due to $Q^{\text{(r)}}$ frequency-dependence, these correlations are different for
each spectrum frequency, showing a rich coherent interaction between the
probe and pump fields. Note that for the spectra of phase noise difference,
nothing happens for $z\ll z_{\rm abs}$, implying that the correlation oscillations shown
in Eq.~\eqref{eq:correso} are qualitatively different from the correlations
implied in Ref.~\cite{rv:richter} and Eq.\eqref{eq:comorichter}.  

For $z\gg z_{\rm abs}$, the correlations reach a constant value given by:
\begin{eqnarray}
{\mathcal S}_c(z,\omega) &=&\frac{1}{2} \big(\cos \left(\theta _1-\theta
_2\right) \left(f_1(\omega )+f_2(\omega )\right)\nonumber\\&&+\cos \left(\theta _1+\theta
_2\right) \left(g_1(\omega )+g_2(\omega )\right)\big)\, .
\end{eqnarray}
Note that, except for initial coherent states ($f_i=g_i=0$), there is always
a formation of correlations.  

We can interpret the generation of the $z>z_{\rm abs}$ constant correlations as
a signature of initial noise in the fields, and the oscillatory correlations
between the fields as a signature that the initial noise spectrum is different
for each field.

 The interchange of noise properties and the oscillatory creation
  and annhilation of correlations can be explained by the fact that the
  two fields are coupled via the atomic
medium.  Due to EIT this coupling is such that the field mean values are unaltered and only the fluctuations get coupled. The atoms base level coherence plays a key role in this
behavior, the oscillatory interchange of noise is complete and the generated correlations reach a
maximum when the
base level coherence of the atoms, $\langle\sigma_{12}\rangle$, is maximal.  When $\langle\sigma_{12}\rangle$ goes to zero 
 there is neither interchange of noise nor generation of correlations between the fields.

When $\alpha_1=\alpha_2$ the oscillatory interchange of noise
properties seems to resemble the noise interchanges that occurs when
two oscillators, with the same frequency, are coupled (see
Ref.~\cite{rv:2oscoupled} for an example with squeezed
states). Differently to the case we are studying, in the two coupled
oscillators the transfer of noise does not depend
on the mean value of the fields: it is always complete. Also in this
case the field mean value is altered during the  interaction.

\section{Carrier frequencies in a detuned two-photon resonance
  ($\delta_1=\delta_2\neq0$)}
\label{sc:detuning}

 In this section we study the case where the carrier frequencies of the pump and probe
field are in a detuned two-photon resonance $\delta=\delta_1=\delta_2\neq 0$. The pump
field is initially in a coherent state and the probe field has initially some general noise.  

We solve equations~\eqref{eq:propagation} with
$\delta_1=\delta_2=\delta$ and $f_1(\omega)=g_1(\omega)=0$ in the
initial conditions Eq.\eqref{eq:ics2} (i.e., a coherent pump field) following
the treatment used in \cite{rv:pablomarc}. In appendix~\ref{ap:2mrescase} we list the analytical expressions for the quadrature noise
spectra of the  pump
($j=1$) and probe ($j=2$) field, and the correlations between them.

\subsection{Asymptotic behavior}
As $Q_\pm^{(i)}$ is always negative, for large propagation lengths,
$z\rightarrow\infty$, we find
\begin{alignat}{1}
  {\mathcal S}_1(z,\omega)\approx&1+({\mathcal
    S}_2(z=0,\omega)-1)\frac{\alpha_1^2
    \alpha_2^2}{(\alpha_1^2+\alpha_2^2)^2}\, ,\label{eq:1inf}\\
  {\mathcal
    S}_2(z,\omega)\approx&1+({\mathcal
    S}_2(z=0,\omega)-1)\frac{\alpha_2^4}{(\alpha_1^2+\alpha_2^2)^2}\, ,\label{eq:2inf}
\end{alignat}
which are reminiscent to similar correlations known from cavity EIT
\cite{rv:agarwal} and the effect of pulse-matching
\cite{rv:harris,rv:fleich1}. The asymptotic behavior does not depend on $\delta$. 
The distance where this asymptotic behavior is dominant is governed by the
exponentials in Eqs.~(\ref{eq:s1}) and ~(\ref{eq:s2}), and is of the order of
\begin{equation}\label{eq:abs} z_{\rm abs}\approx
  1/({\rm Max}(|Q_+^{(i)}|,|Q_-^{(i)}|))\, .
\end{equation}
When $\alpha_1\neq 0$, the asymptotic quadrature fluctuations of the pump field
shows a fraction of the noise properties of the probe field. Also the noise
of the probe field diminishes if ${\mathcal S}_2(z=0,\omega)>1$ and increases otherwise.

For equal field intensities $\alpha_1=\alpha_2$, both
fields have asymptotically the same fluctuations.


\subsection{The probe field is a vacuum squeezed state}
\label{sc:sqdet}
For a broadband vacuum squeezed state ($\alpha_2=0$)the initial conditions
\eqref{eq:ics2} reduce to
   
\begin{eqnarray}
  {\mathcal S}_1(z=0,\omega)&=&1\, ,\\
  {\mathcal S}_2(z=0,\omega)& =&e^{-2\xi}\cos^2\theta+e^{2\xi}\sin^2\theta\, ,
\label{eq:icsvacsq}
\end{eqnarray}
where $\xi$ is a real number characterizing the amount of squeezing.

From Eq. (\ref{eq:s1}) can be concluded that
the pump field quadratures remains as in the initial condition,
$\mathcal{S}_1(\theta,\omega)=1$, as the field propagates.  We also know that
in the case where both fields are in resonance the medium is transparent for
the squeezed vacuum, except for the expected absorption for frequencies detuned
from resonance (see section \ref{sc:resonance} and Ref. \cite{rv:dantan3}).
Nevertheless, when the fields are detuned but in two-photon resonance, although there is no change for
the mean values, the vacuum squeezed state is altered. We can see this by
substituting $\alpha_2=0$ in Eq. (\ref{eq:s2}):
\begin{eqnarray}\label{eq:s2vsdis}
  \mathcal{S}_2(z,\omega)&=&1+\left(e^{2 Q_-^{(i)}z}+e^{2 Q_+^{(i)}z}\right)
  \sinh ^2(\xi )\nonumber\\&&-e^{Q_-^{(i)}z+Q_+^{(i)}z} \cos
  (Q_-^{(r)}z-Q_+^{(r)}z+2 \theta ) \sinh (2 \xi )\, .\nonumber\\
\end{eqnarray}

The propagation of a squeezed vacuum in an EIT medium for the case of detuned
two-photon resonance is characterized by a frequency dependent quadrature
rotation and absorption, characterized by $Q_\pm^{(r)}$ and $Q_\pm^{(i)}$,
respectively. The spectrum absorption of the squeezing is thus different from
the spectrum absorption of the mean value.

\subsubsection{Quadrature rotation}
To better understand the behavior of the coherent part (i.e., neglecting the
absorption part) of the propagation, let us
suppose that $z\ll z_{\rm abs}$. In that case we can replace the
exponentials in Eq.~\eqref{eq:s2} with $1$. Comparing the result with the incoming field, Eq.
(\ref{eq:icsvacsq}), we can write:
\begin{equation}\label{eq:s2vsrot}
  \mathcal{S}_2(z,\omega,\theta=0)=\mathcal{S}_2(z=0,\omega,\theta=(Q_-^{(r)}-Q_+^{(r)})z/2)\, .
\end{equation}
This last equation shows that the propagation of a squeezed probe vacuum in EIT, is equivalent to the rotation of the angle of maximum
squeezing. The velocity of this rotation is given by $Q^{(r)}_\pm$ and is a
function of the detuned two-photon
resonance parameter $\delta$  and spectrum frequency $\omega$.
Due to the dependence of the velocity of
rotation on $\omega$, after propagation the maximum squeezed
probe quadrature would be different for each spectrum frequency. Note that
this behavior can not be modeled using Eq.\eqref{eq:naiveapp}. 
Since the $\theta=0$
quadrature fluctuation of the propagated field is different for each frequency,
we can say that the media would not be transparent for the incoming broadband
$\theta=0$ vacuum squeezed state. 

Note that this result is qualitatively
  different from the known case \cite{rv:fleischstwo-photon}, where as a result
  of the pump detuning all the quadratures are rotated the same amount,
  independently of $\omega$. The difference between our results and the
  results in \cite{rv:fleischstwo-photon} are due to the approximations
  used. In particular we do not
  make any adiabatic approximation. Note also that the frequency-dependence on
  the velocity of quadrature rotation cannot be explained only by a global phase rotation due
  to slow light propagation. 


\subsubsection{Spectrum of squeezing absorption}
\begin{figure}
  \subfigure{\includegraphics[width=6.5cm]{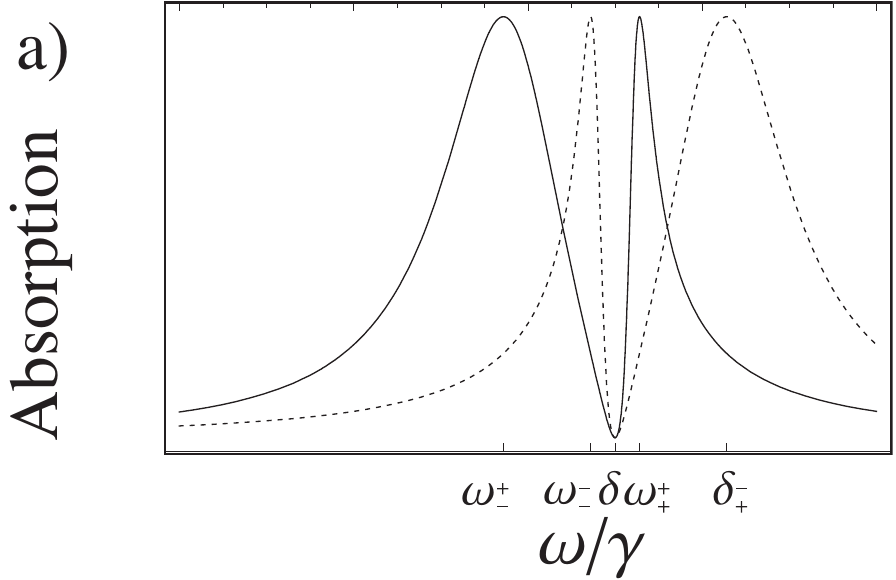}\label{fig:abs_delta_a}}\\
  \subfigure{\includegraphics[width=6.5cm]{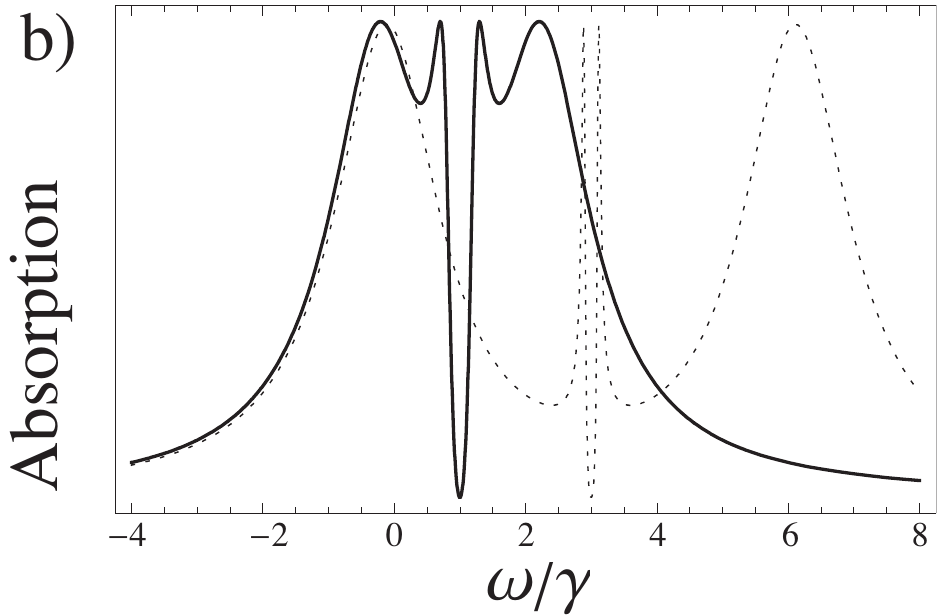}\label{fig:abs_delta_b}}
  \caption{\label{fig:abs_delta} Absorption spectrum for squeezed states for
    the case where there is a detuned two-photon resonance. a) $|Q_+^{(i)}|$ (solid
    line), $|Q_-^{(i)}|$ (dotted line). b) $|Q_+^{(i)}+Q_-^{(i)}|$ for
    $\delta=1\gamma$ (solid line), $\delta=3\gamma$ (dotted line). Note how five
    regions of transparency start to appear in figure b), see text for
    details. Parameter: $\Omega=0.6 \gamma$.}
\end{figure}

The effect of the atom decay rate is, as the field propagates, to dampen the
squeezing of the quadrature with maximum squeezing of the probe
field. For the resonance case, the absorption of the maximum squeezed quadrature follows the
classical absorption spectrum. This is not the case  when there is two
photon resonance but each field is detuned from the atomic transition.
In Eq. \eqref{eq:s2vsdis} it can be observed that the absorption is
proportional to 
$|Q_\pm^{(i)}|$ and the sum $|Q_+^{(i)}+Q_-^{(i)}|$. The two maxima $\omega_\pm$
of each of the $Q_\pm$ are located at
\begin{equation}
\omega^\pm_\pm=\delta+\frac{1}{2}(\pm\delta\pm\sqrt{\delta^2+4\Omega^2})\, .
\end{equation}
Figure~\ref{fig:abs_delta}
shows the absorption spectrum. The absorption spectrum of the noise
differs qualitatively 
from  the absorption spectrum of the mean values Fig.~\ref{fig:res}, only for 
$\omega=\delta$ there is no absorption in either case. We denote the region
around $\omega=\delta$ where absorption is negligible as
$\Delta\omega_\delta$. The length of this spectral region,
$\Delta\omega_\delta$, decreases as
$\delta$ increases.  When $\delta$ increases, the difference between
$\omega^+_+-\omega_-^+=\omega^-_+-\omega_-^-$ increases. When this difference
is much bigger than $\gamma$, a spectral region appears between $\omega^+_+$ and $\omega_+^-$,
and between $\omega^-_-$ and $\omega_-^+$ where absorption is negligible. This
is shown as the dotted line in Fig.~\ref{fig:abs_delta_b}. Then, when
$\delta\gg\gamma,\Omega$, instead of the three regions of transparency
that appears in the spectrum absorption for the
mean values (see Fig.~\ref{fig:res}), there are five regions of transparency: (i) around
$\omega=\delta$;  (ii)
between $\omega^+_-$ and $\omega_-^-$; (iii) between $\omega^-_+$ and
$\omega_+^+$; (iv) when $\omega\gg\omega_+^-$ and (v) $\omega\ll\omega_-^+$. The width of the transparency window for case (i) is of order
$\Delta\omega_\delta\approx 2\Omega^2/\delta$, for case (ii) and (iii) is of order
$\delta$.

Note that
in Ref. \cite[section II-C]{rv:dantan4} , it is concluded that maximum absorption takes place at
$\omega=\delta$, and that there is an absorption of squeezed
states around $\omega=\delta$, and transparency elsewhere. This contradiction
between the results presented here and those from the reference is due to
the fact that they make the $\delta\gg \gamma$ approximation before solving
the equation. As a result of that approximation, they do not obtain the
transparency region around $\omega=\delta$.

\begin{figure}[!ht]
  \includegraphics[width=3in]{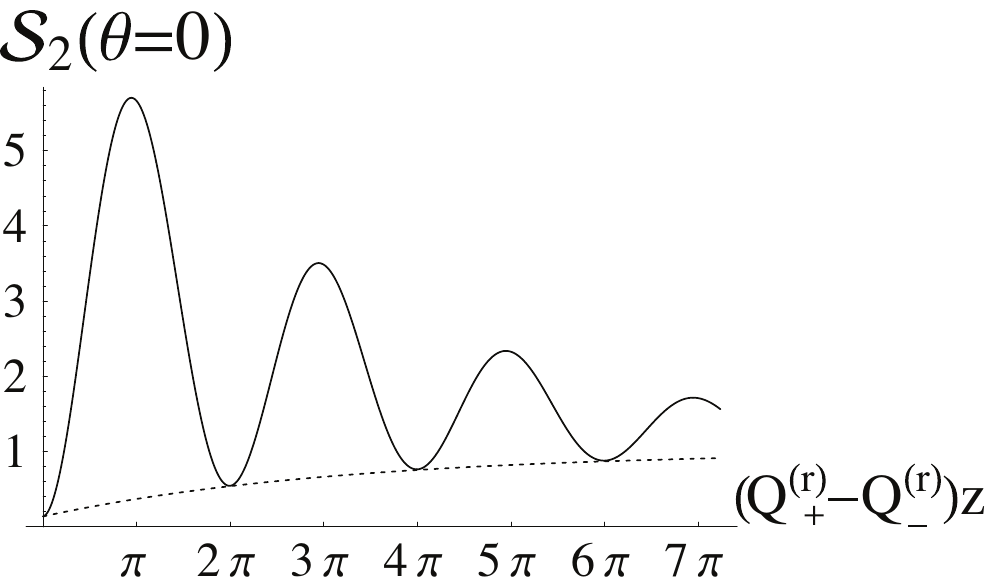}
  \caption{\label{fig:sqvac} Propagation of the $\theta=0$ quadrature
    spectrum of the probe field, initially (at $z=0)$ in a broadband vacuum squeezed state
    with squeezing parameter $\xi=1$, along the $z$-direction in the
    case of a pump detuning such that $Q_+^{(r)}/Q_-^{(r)}=1.106$ (solid
    line), and no pump detuning $Q_+^{(r)}/Q_-^{(r)}=1$ (dashed line). The
    length scale is given by the parameters of the solid line curve.
    Parameters: $Q_+^{(i)}/Q_-^{(i)}=1.224$, $|Q_+^{(r)}/Q_+^{(i)}|=60\pi$
    (solid line), $Q_+^{(i)}/Q_-^{(i)}=1$, $|Q_+^{(r)}/Q_+^{(i)}|=63\pi$
    (dashed line). }
\end{figure}

An example of the interplay between the rotation of the angle of maximum
squeezing and the absorption of the initial squeezing due to the linewidth of
the excited level, given by Eq. (\ref{eq:s2vsdis}), is plotted in Fig.~\ref{fig:sqvac}. 

\subsection{The probe field is a squeezed state with $\alpha_1=\alpha_2$}
\begin{figure}
  \subfigure{\includegraphics[width=6.5cm]{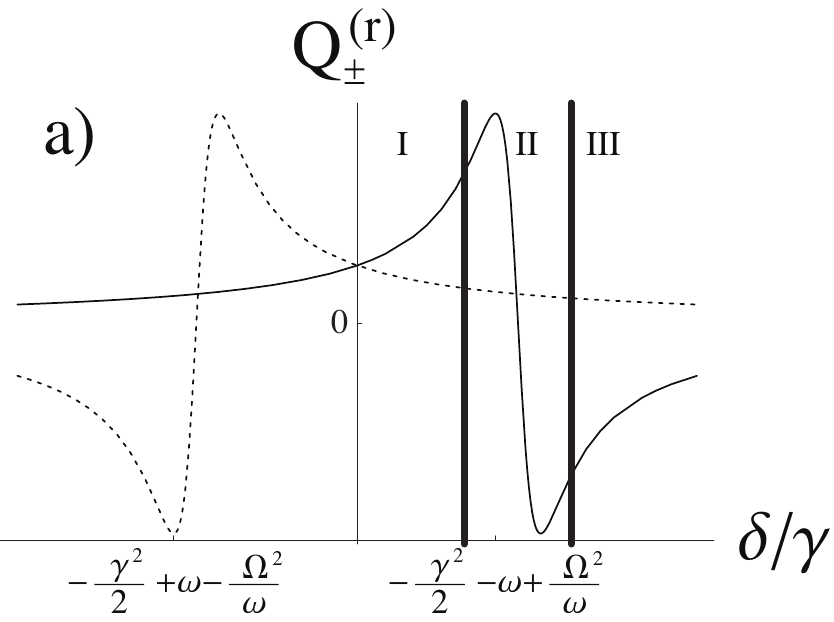}\label{fig:qmasmenosr}}\\
  \subfigure{\includegraphics[width=6.5cm]{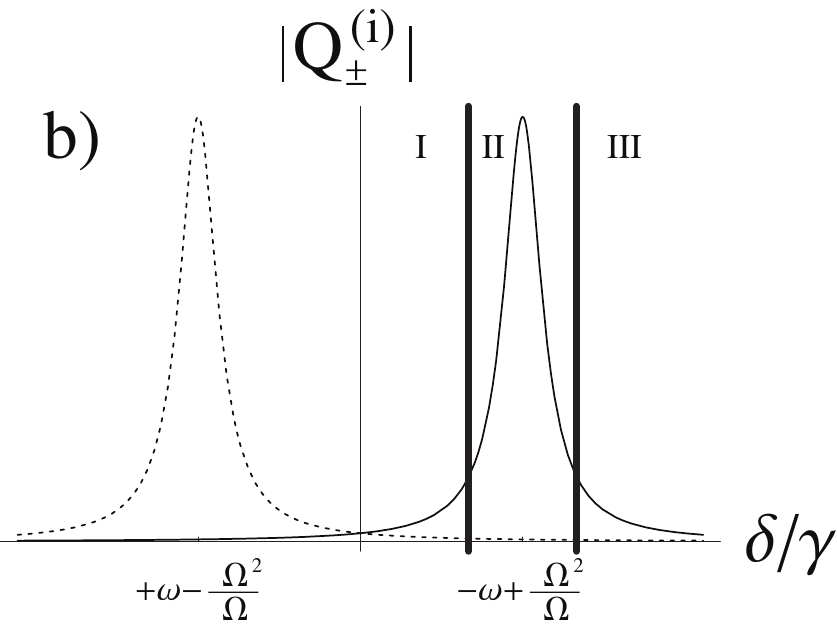}\label{fig:qmasmenosi}}
  \caption{\label{fig:qmasmenos} The oscillating parts of the quadrature
    fluctuation spectrum are proportional to $Q_\pm^{(r)}$. In a) we show
    $Q_+^{(r)}$ (solid line) and $Q_-^{(r)}$ (dashed line) as a function of
    $\delta$. The extrema are obtained for $\delta_{\pm}=-\frac{\gamma
    }{2}\pm\frac{\Omega ^2-\omega ^2}{\omega }$.  The absorbing parts of the
    quadrature fluctuation spectrum oscillations are proportional to
    $Q_\pm^{(i)}$. In b) we show $Q_+^{(i)}$ (solid line) and $Q_-^{(i)}$
    (dashed line) as a function of $\delta$. For $Q_\pm^{(i)}$ the extrema are
    obtained for $\delta_{\pm}=\pm\frac{\Omega ^2-\omega ^2}{\omega }$.
    The maximum of $Q_+^{(i)}$ ($Q_-^{(i)}$) is centered between
    the two extrema of $Q_+^{(r)}$ ($Q_-^{(r)}$).}
\end{figure}

We discuss now the propagation of a squeezed probe field in the case where
$\alpha_1=\alpha_2$.  We separate this discussion into three parts. From the
form of the $Q_+^{(r,i)}$, Eq.~\eqref{eq:qmasmenos}, we can identify three
regions of  qualitatively different behavior, denoted by the Roman numbers I,
II, and III in Fig.~\ref{fig:qmasmenos}. We will therefore divide the study of
the propagation of squeezed fields in EIT media into these three parts for the
case $\delta>0$, the other case being symmetric.

\subsubsection{Near-resonant case (Region I)}\label{ssc:sr}
We analyze here the propagation of the field in the near-resonant case, $|\omega\delta|\ll\Omega^2$.
A qualitatively different behavior from the case $\delta=0$ appears.

\begin{center}
  \begin{figure*}[!ht]
    \includegraphics[width=7in]{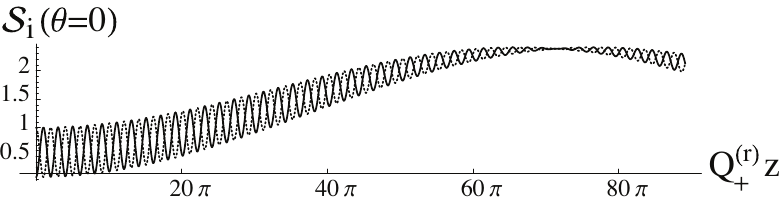}
    \caption{\label{fig:quadratura0zoom} Quantum properties of probe and pump fields
    propagating through an EIT medium with a detuned two-photon resonance. It is shown
    how the fluctuation spectrum of the probe field, initially (at $z=0)$ in a
    squeezed state with squeezing parameter $\xi=1$ propagates along the
    $z$-direction in the case where $z\ll z_{abs}$. The oscillatory transfer of squeezing between probe and pump
      can be clearly seen to be transformed into excess noise until both
      fields reach the same value.  Parameters: $Q_+^{(i)}=Q_-^{(i)}=0$, $Q_+^{(r)}/Q_-^{(r)}=1.014$, $Q_+^{(i)}/Q_-^{(i)}=1.029$}
  \end{figure*}
\end{center}

\begin{figure}[!ht]
  \subfigure{\label{fig:quadratura0a}
    \includegraphics[width=1.6in]{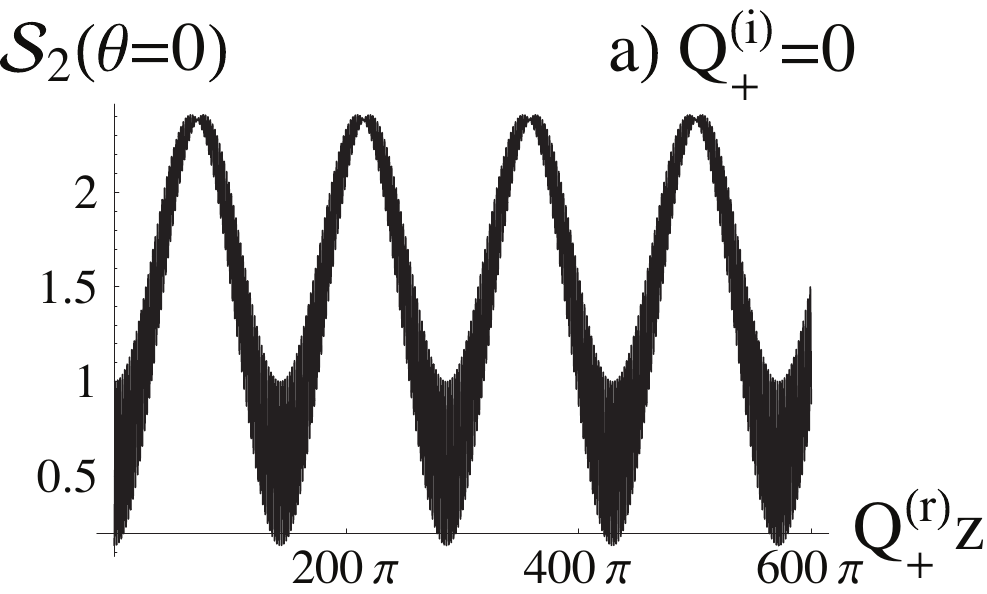}} \subfigure{
    \includegraphics[width=1.6in]{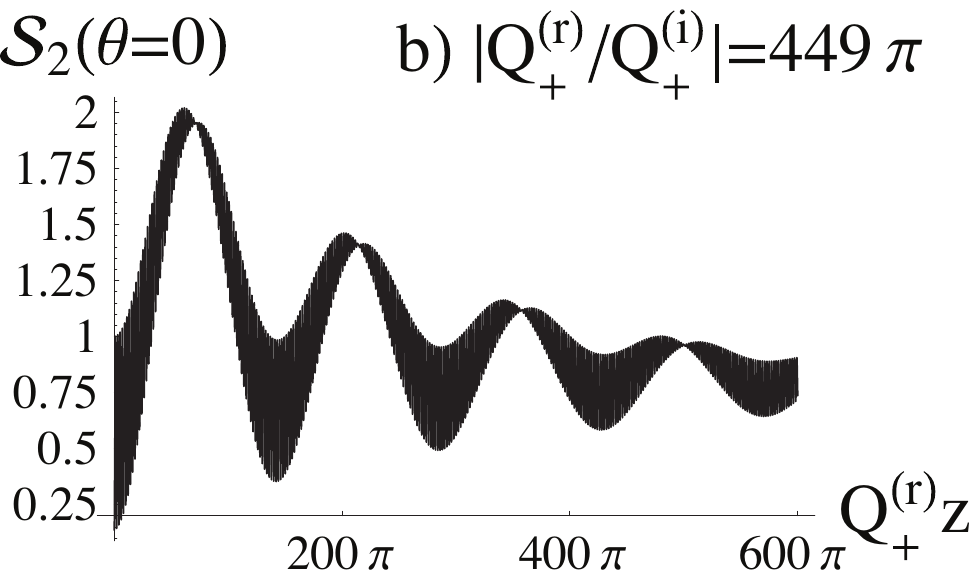}}\\
  \subfigure{
    \includegraphics[width=1.6in]{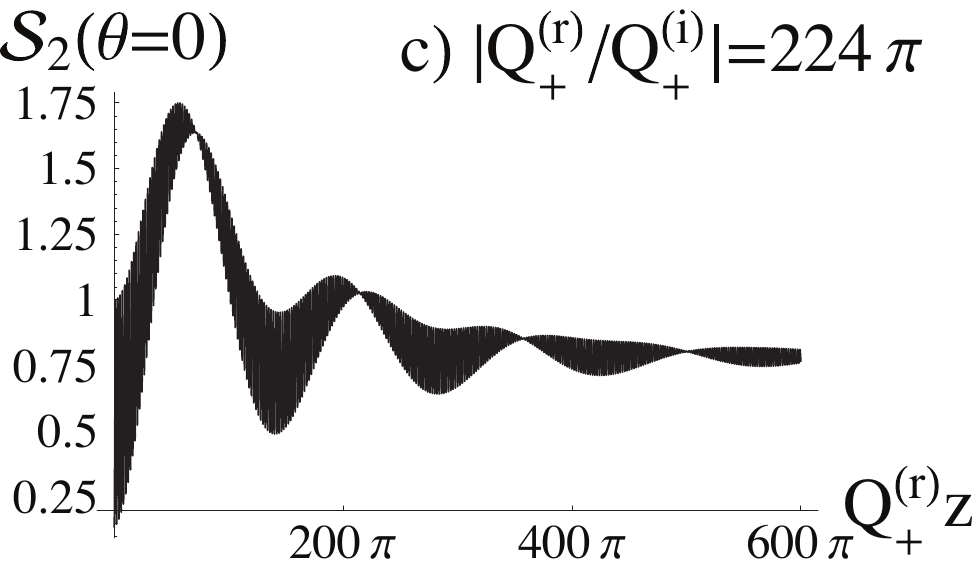}} \subfigure{
    \includegraphics[width=1.6in]{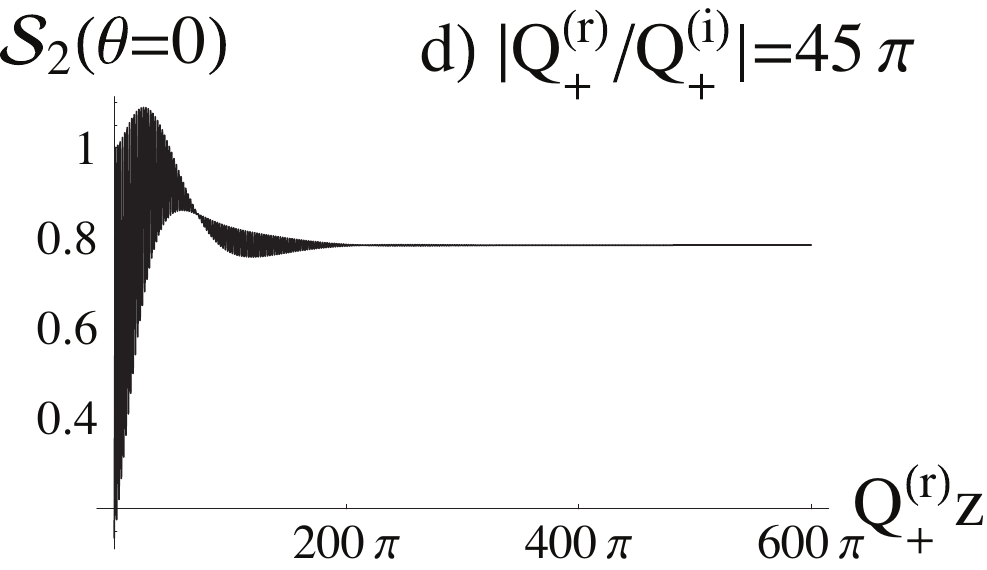}}
  \caption{\label{fig:quadratura0} Quantum properties of probe and pump fields
    propagating through an EIT medium with a detuned two-photon resonance. It is shown
    how the fluctuation spectrum of the probe field, initially (at $z=0)$ in a
    squeezed state with squeezing parameter $\xi=1$ propagates along the
    $z$-direction for different decay rates. The black
area of the plots represents the fast oscillations. a) No decay rate ($Q_+^{(i)}=0$),
    b)-c) $|Q_+^{(r)}/Q_+^{(i)}|=449\pi,224\pi,45\pi$ ($Q_+^{(i)}$ is
    proportional to the decay rate).  The $z_{\rm int}$ scale from minimum
    $e^{-2\xi}$ to $e^{2\xi}/4$ and back can be clearly seen in a).  From b)
    to c) the effect of the absorption scale can be clearly seen. Parameters:
    $Q_+^{(r)}/Q_-^{(r)}=1.014$, $Q_+^{(i)}/Q_-^{(i)}=1.029$,
    $\alpha_1=\alpha_2$.}
\end{figure}

Figs.~\ref{fig:quadratura0zoom} and~\ref{fig:quadratura0a} show a typical example of how the $\theta=0$
quadrature of the field propagates, according to Eqs.~\eqref{eq:s1},\eqref{eq:s2},
for the case of negligible absorption. In Fig.
\ref{fig:quadratura0} we show the effect of the decay rate $\gamma$. The black
area in the plots represents the fast oscillations. There are
three scales that can be seen in Figs.~\ref{fig:quadratura0zoom}
and~\ref{fig:quadratura0}: (i) a quick scale $z_{\rm osc}$, where we have
oscillating transfer of squeezing between pump and probe fields for some regions, (ii) an intermediate oscillating scale,
$z_{\rm int}$, given by the oscillation of the envelope of the fast
oscillations; and (iii) the absorption scale $z_{\rm abs}$ which damps the oscillation
behavior, given by Eq.(\ref{eq:abs}). On the intermediate scale, the probe $\theta=0$ quadrature goes from a squeezed value to a
maximum excess noise ($\mathcal{S}_2(\theta=0)>1$) and then back to the
squeezed value. It is not difficult to show that for $e^{2\xi}\gg 1$ the
maximum of the $\theta=0$ probe quadrature is $\mathcal{S}_2(\theta=0,z_{\rm
  max})\approx e^{2\xi}/4$ and the minimum is $e^{-2\xi}$. 

We proceed now to obtain analytical
expressions for the scales. When $\Omega^2>|\omega(\omega\pm\delta)|$ and $Q_-^{(r)},Q_+^{(r)}>0$, we can
extract fast oscillating terms (given by $Q_-^{(r)},Q_+^{(r)}$) and slow oscillating
terms (given by $Q_-^{(r)}-Q_+^{(r)}$) from Eqs.(\ref{eq:s1}) and
(\ref{eq:s2}). $z_{\rm osc}$ is given by the fast oscillating terms: when
$\omega\delta\ll\Omega^2$, 
\begin{equation}
z_{\rm osc}\approx \pi/Q_-^{(r)}\approx
\pi/Q_+^{(r)}\, . 
\end{equation}
The intermediate oscillating scale, $z_{\rm int}$, is given by
the slow oscillating terms: 
\begin{equation}
z_{\rm int}=2\pi/(Q_-^{(r)}-Q_+^{(r)})\, .
\end{equation}

To gain more insight into the propagation of an initially $\theta=0$
quadrature squeezed state in EIT media, it is necessary to study all the
quadratures.
Excess noise in the $\theta=0$
quadrature after some propagation does not necessarily imply that the state is
no longer a squeezed state. The squeezed
quadrature could have been rotated (see section ~\ref{sc:sqdet}).  
It is worth then,
instead of studying the propagation of a specific quadrature, to study the
propagation of the quadrature where the minimal and maximal fluctuations are
achieved. To calculate these quadratures we minimize Eqs.  (\ref{eq:s1}) and
(\ref{eq:s2}) for $\theta$. We will assume that $z\ll z_{\rm abs}$. Simple
expressions can be obtained for the angle $\theta_{\rm min}$ of the probe
quadrature where the minimum value is reached and the angle $\theta_{\rm max}$
where the maximum value is reached:
\begin{eqnarray}
  \theta_{\rm min,n}&=&\pi  n+\frac{(Q_+^{(r)}-Q_-^{(r)})z}{4}\, ,\label{eq:rotsq}\\
  \theta_{\rm max,n}&=&\theta_{\rm min,n}+\pi/2\, ,
\end{eqnarray}
where $n$ is an integer. Similarly to the case where the probe field is a
vacuum squeezed state (see section~\ref{sc:sqdet}), the quadrature of minimal fluctuations rotates
with propagation.

\begin{figure}[!ht]
    \includegraphics[width=3.5in]{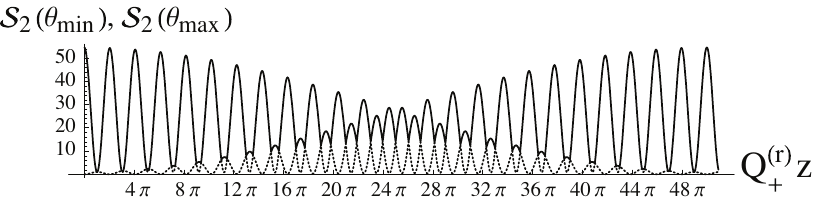}
  \caption{\label{fig:quadraturamaxmin} Propagation of the probe's $\theta_{\rm max}$ quadrature spectrum compared with the probe's $\theta_{\rm min}$ quadrature spectrum for $z\ll z_{\rm
      abs}$.  Here
    $\theta_{\rm max}$ maximize the quadrature and $\theta_{\rm min}$ minimize
    the quadrature.  It is clearly seen that there are
    distances where the $\theta_{\rm max}$ and $\theta_{\rm min}$ quadratures are
    equal. Parameters: $Q_+^{(r)}/Q_-^{(r)}=1.042$,$Q_+^{(i)}=Q_-^{(i)}=0$,
    $\alpha_2=\alpha_1$, $\xi=2$}
\end{figure}

In Fig.~\ref{fig:quadraturamaxmin} we compare the $\theta_{\rm min}$
quadrature with the $\theta_{\rm max}$ quadrature. We observe that for some
values of $Q_+^{(r)} z$, the minimum quadrature and the maximum quadrature of
the probe field reach the same value. This means that for these particular
values of $Q_+^{(r)} z$, all the quadratures have the same value. These values
of $Q_+^{(r)} z$ can be calculated analytically by minimizing
Eqs. (\ref{eq:s1}) and (\ref{eq:s2}) with respect to $z$ such that they are independent of $\theta$. We
obtain
\begin{eqnarray}\label{eq:qzp}
  Q_+^{(r)}z_{2,+}(m)=4 \pi  m+\pi\, , \nonumber\\
  Q_-^{(r)}z_{2,-}(m)=4 \pi  m+\pi\, ,
\end{eqnarray}
where $m$ is an integer.  

The previous results tell us that for an initially squeezed probe field, the propagation has the effect of distributing the noise
between the quadratures
until, for distances defined by Eq. \eqref{eq:qzp}, the noise is equally
distributed between all the quadratures. Then, as the field continues to propagate,
the inverse process starts until a particular mode quadrature is squeezed.
This process repeats until the initial condition of the mode (a $\theta=0$
quadrature squeezed state for the probe field, a coherent state for the pump field) is reached. We understand this dynamics as a
  combination of the rotation of the quadrature of minimal fluctuations due to
  detuning (see
  Eq. \eqref{eq:rotsq})  and the
  interchange of noise between the probe and pump fields characterized by $z_{osc}$.

A similar process occurs for the pump field but with the distances
$z_{1,\pm}(m)$ where all the
quadratures have the same value given by,
\begin{eqnarray}\label{eq:qzb}
  Q_+^{(r)}z_{1,+}(m)=4 \pi  m\, , \nonumber\\
  Q_-^{(r)}z_{1,-}(m)=4 \pi  m\, .
\end{eqnarray}

The value of all the quadratures for these distances oscillates and is given by,
\begin{eqnarray*}
\mathcal{S}_1(z_{1,\pm}(m))&=&\mathcal{S}_2(z_{2,\pm}(m))=\\&& 1+\frac{1}{2} (\cos (Q_\mp^{(r)}z_{2,\pm}(m))+1)
\sinh ^2(\xi )\, .
\end{eqnarray*}

\subsubsection{Intermediate detuned two-photon resonance (region II)}
In region II (see Fig. \ref{fig:qmasmenosr}), it is not difficult to see that
exponential absorption, in the vicinity of the maximum of $Q_+^{(i)}(\delta)$,
makes all the properties
of propagation described in the previous section fade away for distances $z<z_{osc},z_{int}$.

\subsubsection{Large detuned two-photon resonance (region III)}
We now study the case of a large detuned two-photon resonance $|\delta\omega|\gg
\Omega^2,\omega^2$ (region III). For that case,
\begin{equation}\label{eq:ldq}
  Q_+^{(r)}\approx -Q_-^{(r)}\approx-\frac{\mathcal{C} \delta \omega
    ^2}{\frac{\gamma ^2 \omega ^2}{4}+\delta ^2 \omega ^2}\, ,
\end{equation}
\begin{equation*}
  Q_+^{(i)}\approx Q_-^{(i)}\approx-\frac{\mathcal{C} \gamma  \omega
    ^2}{2 \left(\frac{\gamma ^2 \omega ^2}{4}+\delta ^2 \omega
      ^2\right)}\, .
\end{equation*}
If $\delta\gg\gamma$ then $|Q_+^{(r)}|\gg |Q_+^{(i)}|$ and a simple
expression can be found from Eqs. (\ref{eq:s2}) and (\ref{eq:s1}) for
$z<z_{\rm abs}$.
\begin{eqnarray*}
  \mathcal{S}_2(z,\omega)&=&\mathcal{S}_1(z+\pi/Q_+^{(r)},\omega)\approx e^{-2 \xi } \cos
  ^4\left(\frac{Q_+^{(r)} z}{2}\right)\\&&+\sin
  ^2\left(\frac{Q_+^{(r)} z}{2}\right)+\frac{1}{4} e^{2 \xi } \sin
  ^2(Q_+^{(r)} z)\, .
\end{eqnarray*}

\begin{center}
  \begin{figure}[!ht]
    \includegraphics[width=3in]{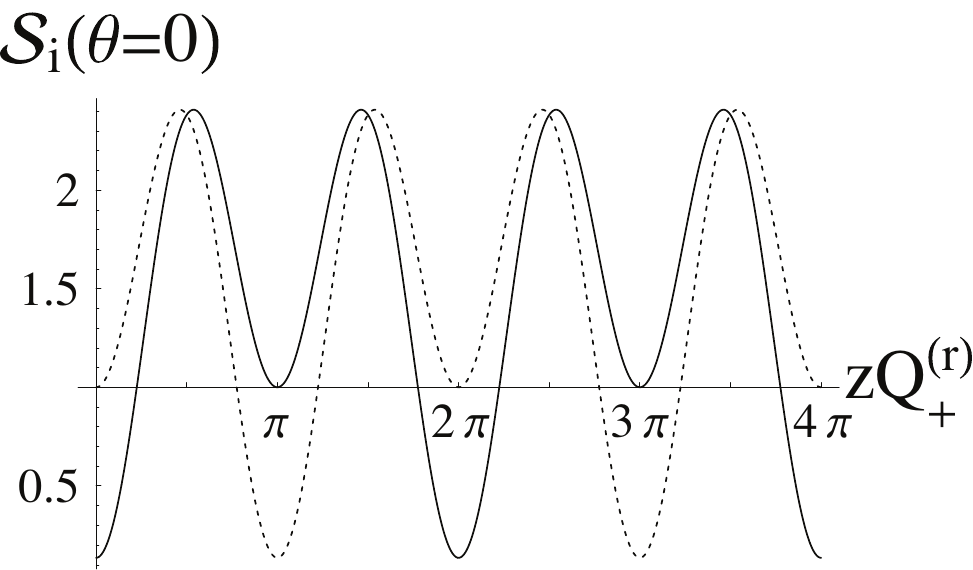}
    \caption{\label{fig:quazerodetuning} The $\theta=0$
      quadrature spectrum of the probe field (solid line) initially (at $z=0$)
      in a squeezed state with squeezing parameter $\xi=1$ and the pump field
      (dashed line) initially in a coherent state, propagates along the
      $z$-direction in the case of large detuning $|\delta|\gg\gamma$,
      $|\delta\omega|\gg \Omega^2,\omega^2$, $\alpha_2=\alpha_1$.}
  \end{figure}
\end{center}

From the former equation, it can be concluded that interchange of quantum
fluctuation between pump and probe happens when $Q_+^{(r)} z=n\pi$, $n$ being an integer.
Differently to the near-resonant case, described in section~\ref{ssc:sr}, the distances where both fields have the same excess
noise is reached before the interchange of noise between the fields occurs. 
In Fig. \ref{fig:quazerodetuning} an example of propagation for large
detuned two-photon resonance is shown. 

\section{Moving atoms and two-photon resonance: Doppler effect}
\label{sc:doppler}

In this section we study numerically the effect of the atoms' Doppler width in
the spectrum of quadrature fluctuation.

As we studied in the previous section, if the pump field is not in resonance
with the dipole transition, then the propagation of the state is hugely altered and
the final propagated state can be very different from the initial state. This
means that for moving atoms, due to the Doppler width, groups
of atoms with different directions and velocities would see different detuned two
photon resonance, and each group would have a different influence on the
propagation of the initial state. Using the superposition principle of the
field and the fact that we are working in a linear approximation, the
propagated final field would be a sum of the propagated field for each group
of atoms $a(t)=\int_{-\infty}^{\infty} \rho(\delta) a_{\delta}(t) d\delta$,
where $a_{\delta}(t)$ is the solution of equation \eqref{eq:fprop} when the
atoms' velocity is such that the detuned two-photon resonance is $\delta$, and
$\rho(\delta)$ is the density of atoms with velocities and directions such that
the detuned two-photon resonance is $\delta$.  The final quadrature spectrum is
then given by
\begin{equation}\label{eq:dopler}
  \mathcal{S}_2(\theta,\omega)_{\Delta\delta}=\int_{-\infty}^{\infty}\int_{-\infty}^{\infty} \rho(\delta_1)\rho(\delta_2) \mathcal{S}_2(\theta,\omega,\delta_1,\delta_2)
  d\delta_1d\delta_2\, ,
\end{equation}
where
\begin{equation} \label{eq:doplerin}
{\mathcal S}(\theta,\omega,\delta_1,\delta_2) =
  \int\limits_{-\infty}^\infty e^{-i\omega t} \langle \delta
  Y_{\delta_1}^\theta(t)\delta Y_{\delta_2}^\theta(0)\rangle dt\, ,
\end{equation}
and $\delta Y_{\delta_i}^\theta(t)=\delta a_{\delta_i}(t) \exp(-i\theta) +
\delta a_{\delta_i}^\dagger(t) \exp(i\theta)$.  We will assume that
$\rho(\delta)$ is a Gaussian with variance $\Delta\delta$.

To obtain the effect of the Doppler width in the quadrature spectrum we
calculate, from Eqs.~\eqref{eq:propagation}, ${\mathcal
  S}(\theta,\omega,\delta_1,\delta_2)$ following
a method similar to the one used in \cite{rv:pablomarc}. Then we
numerically integrate Eq.~(\ref{eq:dopler}).

The frequency $\omega$ obtained from Fourier transforming Eqs.~(2) is the
spectrum frequency measured from the field carrier frequencies. For fixed
$\omega$, changing $\delta$ also changes the spectrum frequency with respect
to the dipole transition of the atom. In order to correctly obtain the Doppler
effect in the quadrature, before integrating Eq. (\ref{eq:dopler}) we
substituted $\omega$ by $\omega-\delta$. After this substitution, and in this
section, $\omega$ represents the spectrum frequency with respect to the atomic
transition.

\subsection{The probe field is squeezed vacuum}
\begin{figure}[ht!]
  \includegraphics[width=3in]{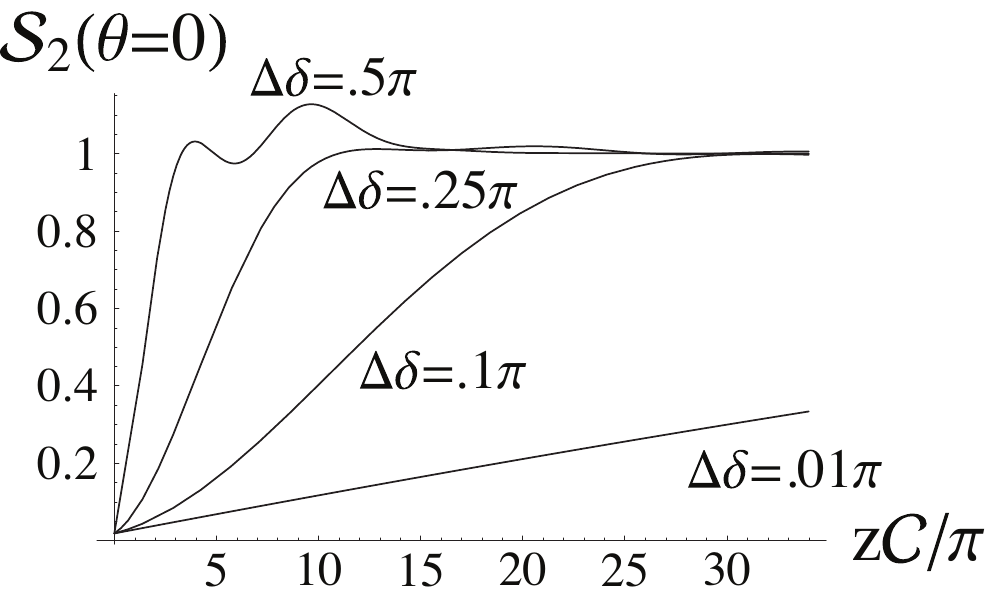}
  \caption{\label{fig:sqvacdop} Propagation of the $\theta=0$ quadrature
    spectrum of the probe field initially (at $z=0)$ in a vacuum squeezed state
    with squeezing parameter $\xi=2$ along the $z$-direction in a
    medium composed of atoms with Doppler width
    $\Delta\delta=.01\gamma,.1\gamma,.25\gamma,.5\gamma$. It is clearly seen
    how the Doppler effect affect the propagation of a vacuum squeezed state.
    Parameters: $\gamma_1=\gamma_2$, $g_1=g_2=\gamma/10$, $\Omega_1=1\gamma$,
    $\Omega_2=0$, $\omega=\gamma/10$,$N=10^{12}$.}
\end{figure}

We now suppose that the probe field is in a broadband squeezed vacuum. This means that $\alpha_2=0$. In this
case, for atoms at rest, and a coherent pump field in resonance with its corresponding transition,
the medium is transparent for the squeezed vacuum (see section
\ref{sc:resonance} and Ref. \cite{rv:dantan3}).  In Fig. \ref{fig:sqvacdop} we
show the numerically calculated probe $\theta=0$ quadrature spectrum for an
initially squeezed vacuum propagating in a medium composed of atoms with
Doppler distribution with width $\Delta \delta$.   
For $\Delta\delta=.01\gamma$, the detuned two-photon resonance due
to the Doppler effect is small and does not affect the propagation. As
$\Delta\delta$ increase, the effect of the different quadrature
rotation on the incoming field, due to atoms with different detuned two-photon
resonance, begins to be important. When
$\Delta\delta\gtrapprox .25\gamma$ we even have excess noise in the
$\theta=0$ quadratures for some distances. The excess noise is larger the
larger $\Delta\delta$. In the limit
$z\rightarrow\infty$ it can be calculated from Eq.(\ref{eq:2inf}) that the
$\theta=0$ quadrature will be $1$. 

\subsection{The probe field is squeezed state}
We study now how the Doppler effect affects the propagation of a squeezed
state as a probe initial condition. We will suppose that the mean value of the
probe field is equal to the mean value of the pump field, $\alpha_1=\alpha_2$.
We will also suppose that the carrier frequencies of both fields are in
resonance with their respective dipole transitions of the atoms at rest.

\begin{figure}[!ht]
  \subfigure{\label{fig:quadopa}
    \includegraphics[width=1.6in]{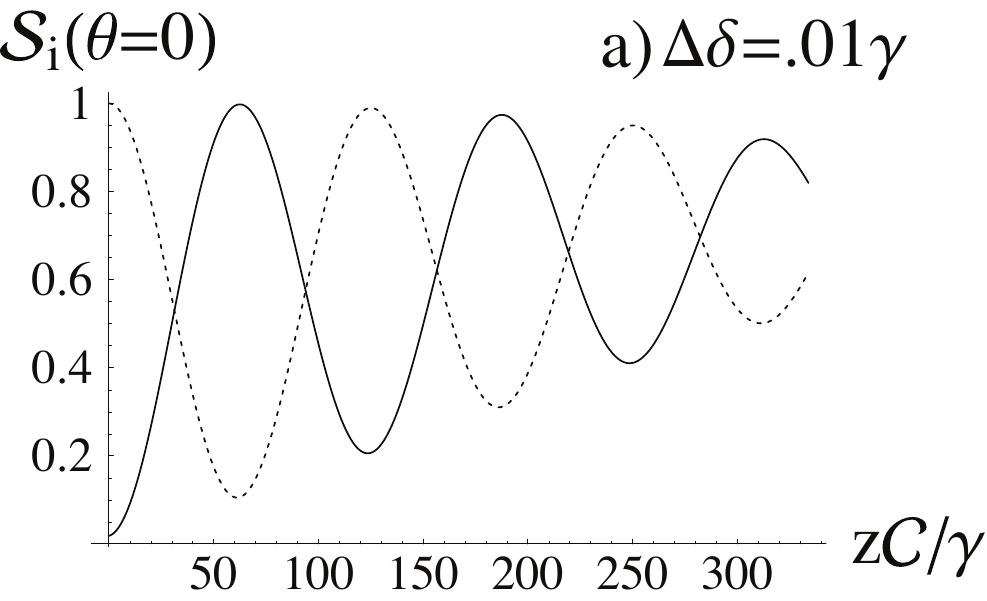}} \subfigure{\label{fig:quadopb}
    \includegraphics[width=1.6in]{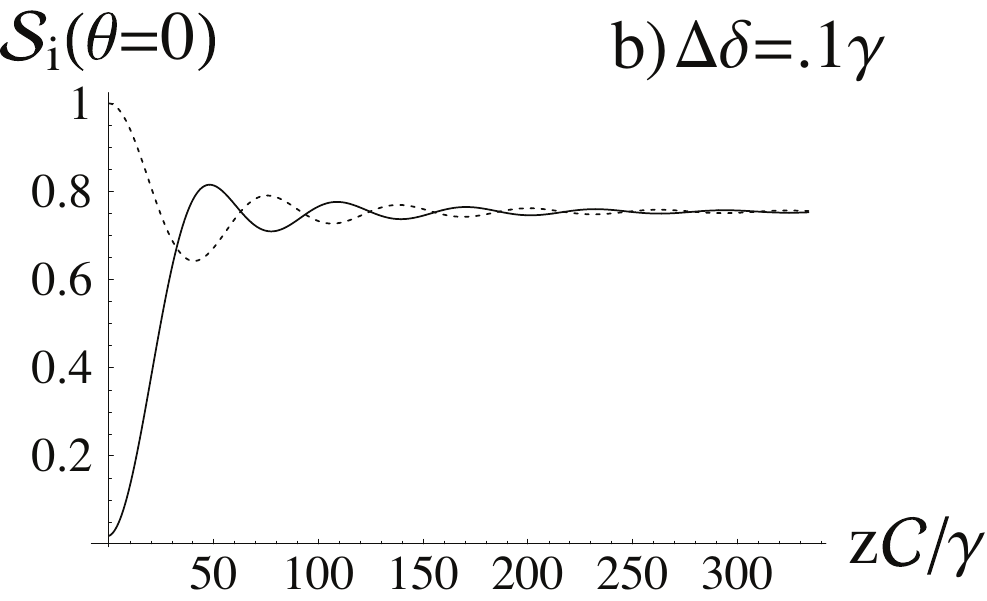}}\\
  \subfigure{\label{fig:quadopc}
    \includegraphics[width=1.6in]{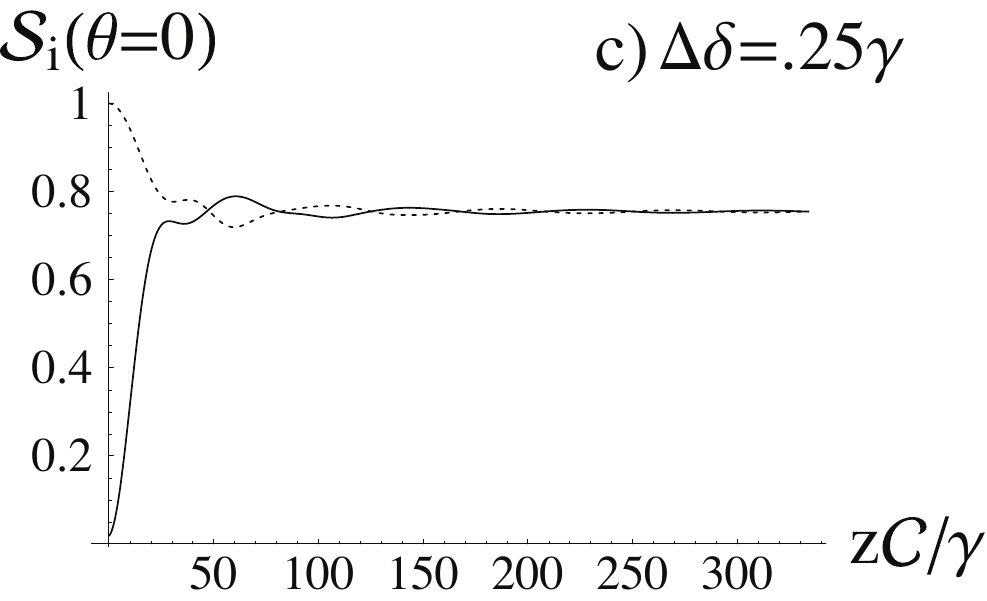}} \subfigure{\label{fig:quadopd}
    \includegraphics[width=1.6in]{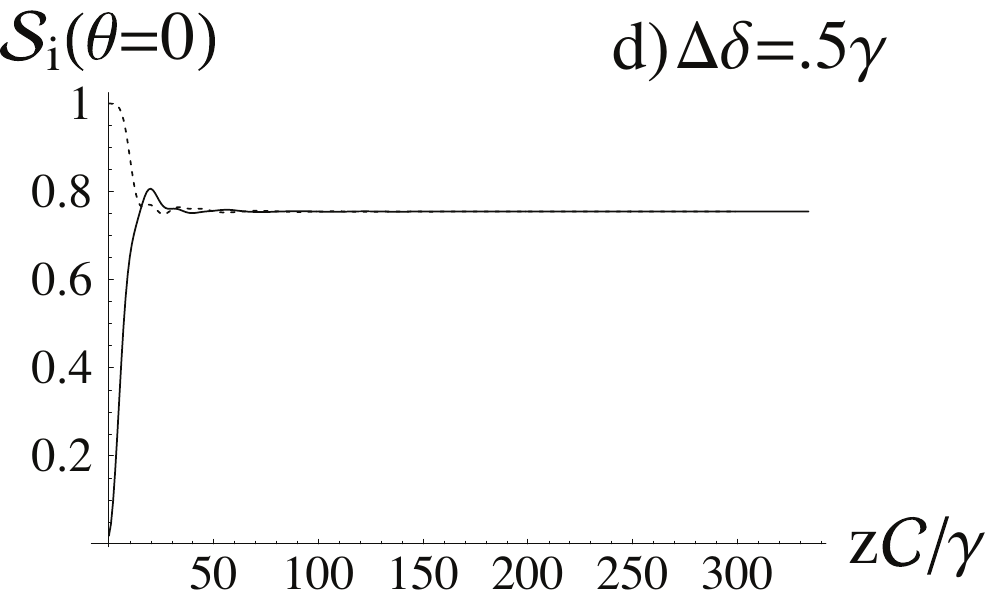}}
  \caption{\label{fig:quadop}Propagation of the $\theta=0$ quadrature
    spectrum of the probe field initially (at $z=0)$ in a squeezed state with
    squeezing parameter $\xi=2$ along the $z$-direction in a medium
    composed of atoms with Doppler width
    $\Delta\delta=.01\gamma,.1\gamma,.25\gamma,.5\gamma$. It is clearly seen
    how the Doppler effect can destroy the oscillatory transfer of squeezing
    between the probe and pump field.  Parameters:$\gamma_1=\gamma_2$,
    $g_1=g_2=\gamma/10$, $\Omega_1=\Omega_2=\gamma$,
    $\omega=\gamma/10$,$N=10^{12}$.}
\end{figure}

Numerical results are presented in Fig. \ref{fig:quadop} for
$g_1=g_2=\gamma/10$, $\Omega_1=\Omega_2=1\gamma$,
$\omega=\gamma/10$,$N=10^{12}$, $\xi=2$. In Fig.\ref{fig:quadopa}, for a
Doppler width of $\Delta\delta=0.01\gamma$, the behavior of the state
propagation is very similar to the studied in section \ref{sc:resonance}. From
Fig.\ref{fig:quadopb} to Fig.\ref{fig:quadopc}, it can be seen how the
increase of the Doppler width starts to destroy the interchange of squeezing
properties between the probe and pump field and the differences between
both fields vanish. In Fig.\ref{fig:quadopd} both fields have practically the same
$\theta=0$ quadrature spectra for $z\mathcal{C}/\gamma>50$.

These results impose restrictions in the allowed atoms' Doppler width for
studying quantum state propagation in EIT media.

\section{Conclusions}

We have shown that in the propagation of pump and probe fields through an EIT
medium, the initial noise properties of the field are not conserved, except for the carrier
frequencies which drive the atoms on two-photon resonance. We found a series
of novel behavior in the propagations.

The results reported in Ref.~\cite{rv:pablomarc} extend to any initial spectrum
of noise. The effect of coherent propagation (i.e., neglecting the exponential
decay terms) is to interchange the noise properties between the probe and pump
field as the field propagates. The interchange is maximal when both fields
have comparable intensities. The frequency of the interchange of noise properties depends on
the spectrum frequency.  This implies that the noise spectrum of the outcoming
field can be completely different to the noise spectrum of the incoming field.  

When the initial noise spectrum is different between the fields, as the field
propagates there is  an
oscillatory creation and annihilation of correlations between the pump and probe fields.

The oscillatory interchange of noise properties between the two fields is reminiscent
to the case of two coupled harmonic oscillators \cite{rv:2oscoupled}. However, in the
case presented here, the coupling is realized by a quantum system, namely the
atomic media, which shows some special properties due to EIT: In terms of mean values,
there is no excited state population, and thus
the fields and the atomic media are uncoupled.  Consequently, the mean values are 
stationary, in contrast to the case of the two oscillators. Coupling is enabled by 
the quantum fluctuations of the atomic dipole moments, and the ground state coherence 
conveys the interaction between the two fields. Indeed, the generated correlations are 
most strongly present when $\langle\sigma_{12}\rangle$ is maximal. Apart from the 
coherent dynamics, the atomic media introduces noise due to the finite lifetime
of the excited state level. This is reflected in the
decay of the oscillatory interchange of noise properties between the fields at a certain time scale.

The effect of a detuned two-photon resonance in the propagation of an
initially broad band vacuum squeezed state as the probe field and a coherent
state as the pump field, is to rotate the quadrature of maximum squeezing as the
field propagates. The velocity of the rotation is a function of the detuned
two-photon resonance
and spectrum frequency. With a frequency-dependent velocity of rotation,
the outcoming field can be completely  different to the incoming field. This
frequency-dependent phase is different to the expected global phase (the same
phase for all frequencies) due to slow velocity propagation inside the
medium. 

When both fields have comparable Rabi frequencies, with the probe field
initially in a squeezed state and the  pump field in a coherent state, the
effect of a detuned two
photon resonance in the propagation can be described as a combination of a rotation of the
maximum squeezed quadrature with an interchange of squeezing properties
between the pump and probe fields. The effect of this combination is to redistribute the noise in such a way that gives rise to mode-dependent
propagation distances where all the quadratures have the same noise. As the
field continues to propagate we recover the mode squeezed state.

Besides the fields' coherent propagation, the squeezing properties of the quadratures
are absorbed as the field propagates. The absorption spectrum for the
squeezing properties is
different to the absorption spectrum for the mean values. It is interesting to
note that the transparency window around spectrum frequency $\omega=\delta$ is
inversely proportional to the detuned two-photon resonance and can be very narrow.  

An important result, is the influence of the atoms' Doppler width on the
propagation of a squeezed probe field. Differently to the propagation of the
mean values, where the Doppler
effect has a small influence, and due to the
considerable effect that detuned two-photon resonance has for states propagating
in an EIT medium, the atoms' Doppler
width can destroys the initial squeezing properties of
the field as it propagates. 

\acknowledgments
PBB gratefully acknowledges support from DGAPA and CONACYT.
MB gratefully acknowledges support from the Alexander-von-Humboldt
foundation. We thank David Sanders for carefully reading the manuscript.

\appendix
\section{Analytical expressions for the noise and correlation spectra.}\label{ap:rescase}
We solve equations~\eqref{eq:propagation} with
$\delta_1=\delta_2=0$ and noise spectrum given by~\eqref{eq:ics2}, following
the treatment used in \cite{rv:pablomarc}. We obtain the following analytical results. 
\subsection{The resonance case}
For the resonance case,
$\delta_1=\delta_2=0$, and noise spectrum given by~\eqref{eq:ics2} we obtain
the following expressions for the pump $\mathcal{S}_1$ and probe
$\mathcal{S}_2$ quadrature noise spectra:
\begin{widetext}
\begin{eqnarray}\label{eq:s1res}
    \mathcal{S}_1(z,\omega)&=&\frac{1}{\left(\alpha _1^2+\alpha
      _2^2\right){}^2}\Bigg\{4 e^{-Q^{\text{(i)}} z} \cos \left(Q^{(r)}
    z\right) \alpha _2^2 \left(f_1(\omega )-f_2(\omega )+\cos (2 \theta )
    \left(g_1(\omega )-g_2(\omega )\right)\right) \alpha _1^2\nonumber\\&&+2 e^{-2 Q^{(i)} z}
   \alpha _2^2 \left(f_2(\omega ) \alpha _1^2+\alpha _2^2 f_1(\omega )+\cos (2
   \theta ) \left(g_2(\omega ) \alpha _1^2+\alpha _2^2 g_1(\omega )\right)\right)\nonumber\\&&+2
   \left(f_1(\omega ) \alpha _1^2+\alpha _2^2 f_2(\omega )+\cos (2 \theta )
   \left(g_1(\omega ) \alpha _1^2+\alpha _2^2 g_2(\omega )\right)\right)
   \alpha _1^2\Bigg\}+1\, ,
\end{eqnarray}

\begin{eqnarray}\label{eq:s2res}
    \mathcal{S}_2(z,\omega)&=&\frac{1}{\left(\alpha _1^2+\alpha _2^2\right){}^2}\Bigg\{4 e^{-Q^{(i)} z} \cos
    \left(Q^{(r)} z\right) \alpha _2^2 \alpha_1^2\left(-f_1(\omega )+f_2(\omega
    )+\cos (2 \theta ) \left(g_2(\omega )-g_1(\omega )\right)\right) \nonumber\\&&+2 e^{-2
   Q^{(i)} z} \alpha _1^2\left(f_2(\omega ) \alpha _1^2+\alpha _2^2 f_1(\omega
    )+\cos (2 \theta ) \left(g_2(\omega ) \alpha _1^2+\alpha _2^2 g_1(\omega
    )\right)\right) \nonumber\\&&+2 \alpha
   _2^2 \left(f_1(\omega ) \alpha _1^2+\alpha _2^2 f_2(\omega )+\cos (2 \theta
    ) \left(g_1(\omega ) \alpha _1^2+\alpha _2^2 g_2(\omega )\right)\right)
    \Bigg\}+1\, .
\end{eqnarray}

For the case, $g_1=g_2$ and $\alpha_1=\alpha_2$ we have for the
correlation spectrum
\begin{eqnarray}\label{eq:cores}
\mathcal{S}_c(z,\omega)&=&\frac{1}{2} \big(\cos \left(\theta _1-\theta
_2\right) (f_1(\omega )+f_2(\omega
))+\cos \left(\theta
   _1+\theta _2\right) \left(g_1(\omega )+g_2(\omega
   )\right)\nonumber\\&&+2
e^{-Q^{\text{(i)}} z} \sin \left(Q^{\text{(r)}} z\right)
   \left(\sin \left(\theta _1-\theta _2\right) \left(f_2(\omega )-f_1(\omega
   )\right)+\sin \left(\theta _1+\theta _2\right) \left(g_1(\omega )-g_2(\omega )\right)\right)\nonumber\\&&-e^{-Q^{\text{(i)}} z} \left(\cos \left(\theta _1-\theta _2\right)
   \left(f_1(\omega )+f_2(\omega )\right)+\cos \left(\theta
   _1+\theta _2\right) \left(g_1(\omega )+g_2(\omega )\right)\right)\big)\, ,
\end{eqnarray}

with 
\begin{subequations}\label{eq:qri}
  \begin{eqnarray}
    Q&=& \omega {\mathcal C} \frac{1}{\Omega^2-\omega^2+i\omega\gamma/2}\, ,\\
    Q^{(r)}\equiv{\rm Re} Q &=& \omega {\mathcal C}\frac{\Omega^2-\omega^2}{\left[\Omega^2-\omega^2\right]^2+\omega^2\gamma^2/4}\, , \\
    Q^{(i)}\equiv{\rm Im} Q&= &{\mathcal C}\frac{-\omega^2\gamma/2}{\left[\Omega^2-\omega^2\right]^2+\omega^2\gamma^2/4}\, ,
  \end{eqnarray}
\end{subequations}
\begin{eqnarray*} {\mathcal C}&\equiv& \frac{N
    (g_1^2\Omega_2^2+g_2^2\Omega_1^2}{\Omega^2 c})\, ,\\
  \Omega&\equiv&\sqrt{\Omega_1^2+\Omega_2^2}\, .
\end{eqnarray*}

\subsection{Detuned two-photon resonance case}\label{ap:2mrescase}
For the detuned two-photon resonance case,
$\delta_1=\delta_2=\delta$, and  assuming that $f_1=g_1=0$ in
Eqs.~\eqref{eq:ics2}, we obtain
\begin{alignat}{1}\label{eq:s1}
    \mathcal{S}_1(z,\omega)=&\frac{{\alpha_1}^2
      {\alpha_2}^2}{\left({\alpha_1}^2+{\alpha_2}^2\right)^2}\Big\{f_2(\omega)\left[2+e^{2 {Q_-^{(i)}}z}+e^{2 {Q_+^{(i)}}z} -2 e^{{Q_-^{(i)}}z}
      \cos {Q_-^{(r)}z}
      -2 e^{{Q_+^{(i)}}z} \cos {Q_+^{(r)}z}\right] \nonumber\\
    &\hspace{1.8cm}+2 g_2(\omega) \Big[\cos 2 \theta
    + e^{{Q_-^{(i)}}z+{Q_+^{(i)}}z} \cos ({Q_-^{(r)}}z-{Q_+^{(r)}}z+2 \theta )\nonumber\\
    &\hspace{4.5cm}-e^{{Q_+^{(i)}}z} \cos ({Q_+^{(r)}}z-2 \theta)
    -e^{{Q_-^{(i)}}z} \cos ({Q_-^{(r)}}z+2 \theta)\Big]\Big\}\, ,
  \end{alignat}
  \begin{alignat}{1}\label{eq:s2}
    \mathcal{S}_2(z,\omega)&=\frac{1}{\left({\alpha_
          1}^2+{\alpha_2}^2\right)^2}\nonumber\\&\Big\{{\alpha_1}^4\left[f_2(\omega)\left(e^{2 {Q_-^{(i)}}z}+e^{2 {Q_+^{(i)}}z}\right) +2g_2(\omega)\;
    e^{{Q_-^{(i)}}z+{Q_+^{(i)}}z} \cos
      ({Q_-^{(r)}}z-{Q_+^{(r)}}z+2 \theta )
    \right] \nonumber\\
    &\hspace{2.3cm}+{\alpha_2}^2 {\alpha_1}^2\Big[2 f_2(\omega)
    \left(e^{{Q_-^{(i)}}z} \cos {Q_-^{(r)}z}+e^{{Q_+^{(i)}}z} \cos
      {Q_+^{(r)}z}\right) \nonumber\\
    &\hspace{3.55cm}+2g_2(\omega) \left(e^{{Q_+^{(i)}}z} \cos
      ({Q_+^{(r)}}z-2 \theta )+e^{{Q_-^{(i)}}z} \cos ({Q_-^{(r)}}z+2
      \theta )\right) \Big] \nonumber\\
    &\hspace{2.3cm}+2{\alpha_2}^4 \left( f_2(\omega) + g_2(\omega)
      \cos 2 \theta\right)\Big\}\, .
  \end{alignat}
For the correlations we have, assuming $\alpha_1=\alpha_2$ and $g_1=g_2$
\begin{alignat}{1}\label{eq:co}
    \mathcal{S}_{c}(z,\omega,\theta_1,\theta_2)&=-\frac{1}{2} e^{Q_-^{(i)}z}
    \left(\cos \left(\theta _1-\theta _2\right)(f_1(\omega )+f_2(\omega ))+2 \sin ({Q_-^{(r)}}z) \sin
    \left(\theta _1-\theta _2\right)(f_1(\omega )-f_2(\omega ))\nonumber\right.\\&\left.+\cos \left({Q_-^{(r)}}z+\theta _1+\theta
    _2\right) \left(g_1(\omega )-g_2(\omega
   )\right)\right)\nonumber\\&+\frac{1}{2} e^{Q_+^{(i)}z} \cos
    \left(Q_+^{(r)}z-\theta _1-\theta _2\right) \left(g_1(\omega )-g_2(\omega
    )\right)\nonumber\\&-\frac{1}{2} e^{Q_-^{(i)}z+Q_+^{(i)}z} \cos
   \left({Q_-^{(r)}}z-Q_+^{(r)}z+\theta _1+\theta _2\right) \left(g_1(\omega
   )+g_2(\omega )\right)\nonumber\\&+\frac{1}{2} \left(\cos \left(\theta
   _1-\theta _2\right) (f_1(\omega )+ f_2(\omega ))+\cos \left(\theta _1+\theta _2\right) \left(g_1(\omega )+g_2(\omega )\right)\right)
\end{alignat}
with
\begin{subequations}\label{eq:qmasmenos}
  \begin{eqnarray}
    Q_\pm &=& \omega {\mathcal C} \frac{1}{\Omega^2-\omega(\omega\pm\delta)+i\omega\gamma/2}\\
    Q_\pm^{(r)}\equiv{\rm Re} Q_\pm &=& \omega {\mathcal C}\frac{\Omega^2-\omega(\omega\pm\delta)}{\left[\Omega^2-\omega(\omega\pm\delta)\right]^2+\omega^2\gamma^2/4}\\
    Q_\pm^{(i)}\equiv{\rm Im} Q_\pm &= &{\mathcal C}\frac{-\omega^2\gamma/2}{\left[\Omega^2-\omega(\omega\pm\delta)\right]^2+\omega^2\gamma^2/4}
  \end{eqnarray}
\end{subequations}
\begin{eqnarray*} {\mathcal C}&\equiv& \frac{N
    (g_1^2\Omega_2^2+g_2^2\Omega_1^2}{\Omega^2 c})\\
  \Omega&\equiv&\sqrt{\Omega_1^2+\Omega_2^2}
\end{eqnarray*}
\end{widetext}
\bibliography{../tesis}

\end{document}